\documentclass[aps,prl,floatfix,twocolumn]{revtex4-1}

\usepackage{epsfig}
\usepackage{verbatim}	 
\usepackage{amssymb}
\usepackage{bm}
\usepackage{amsmath}
\usepackage{psfrag}
\usepackage{color}
\usepackage[latin1]{inputenc}
\usepackage[english]{babel}
\usepackage{graphicx}
\usepackage{amscd}
\usepackage{eucal}
\usepackage{hyperref}
\usepackage{lipsum}
\usepackage{mathtools}
\usepackage{multirow}
\usepackage{natbib}

\newcommand{\Kf}{K\hspace{-0.8mm}f}

\begin{document}

\title{Noise-Induced Desynchronization and Stochastic Escape from Equilibrium in  Complex Networks}

\author{M.~Tyloo\textsuperscript{1,4}, R.~Delabays\textsuperscript{2,4}, and Ph.~Jacquod\textsuperscript{3,4}}
\affiliation{\textsuperscript{1} Institute of Physics, \'Ecole Polytechnique F\'ed\'erale de Lausanne, CH-1015 Lausanne, Switzerland. \\
\textsuperscript{2} Automatic Control Laboratory, Swiss Federal Institute of Technology, CH-8092 Z\"urich,  Switzerland.\\
\textsuperscript{3} Department of Quantum Matter Physics, University of Geneva, CH-1211 Geneva, Switzerland\\
\textsuperscript{4} School of Engineering, University of Applied Sciences of Western Switzerland HES-SO, CH-1951 Sion, Switzerland. }

\date{\today}

\begin{abstract}
Complex physical systems are unavoidably subjected to external environments
not accounted for in the set of differential equations that models them. The resulting perturbations are standardly 
represented by noise terms. We derive conditions under which such noise terms perturb the dynamics strongly
enough that they lead to stochastic escape from the initial basin of attraction of an initial stable equilibrium state
of the unperturbed system. Focusing on Kuramoto-like models
we find in particular that, quite counterintuitively, systems with 
inertia leave their initial basin faster than or at the same time as systems without inertia, 
except for strong white-noise perturbations.
\end{abstract}

\maketitle

\textbf{Introduction.}	
Complex physical systems are mathematically modelled as dynamical systems. Equilibrium states, if they exist,
are determined and characterized by fixed points, limit cycles and tori, or even strange attractors of the corresponding
differential equations~\cite{Ott02}. In principle the latter should be complemented
by stochastic terms to account for unavoidable perturbations from unaccountable environmental degrees of freedom~\cite{Kam76}.
A central question of broad interest is to determine the magnitude and statistical properties of the relevant stochastic terms
that could lead to the loss of equilibrium or induce transitions between different equilibria. Some physically 
important situations where such 
stochastic escape phenomena may occur are electric power grids with high penetration of fluctuating 
renewable energy sources~\cite{Mac08,Aue17,Scha17}, 
superconducting rings~\cite{Gou87} and Josephson junction arrays~\cite{Ili08} subjected to noisy magnetic fields, as well as
neuronal systems subjected to synaptic, ion-channel, neurotransmitter or membrane potential noise~\cite{Bra94,Liu18}. 

Despite decades of investigations, theoretical studies of problems related to stochastic escape 
are generally extensions of the pioneering work of Kramers~\cite{Kra40},
which relates chemical reaction rates to action integrals between different potential minima. 
The problem is analytically tractable in low dimensions only, and
several recent works considered noise-induced large fluctuations in the 
dynamical behavior of higher-dimensional network-coupled systems
through the numerical determination of action minimizing paths~\cite{Dev12,Scha17,Hin16,Hin18}. A better 
analytical understanding of the interplay of noise characteristics with the network topology is clearly desirable.
In this manuscript we propose a resolutely different approach to stochastic escape
from stable equilibria in complex, network-coupled dynamical systems, incorporating noise characteristics as well 
as network dynamics and topology.

\begin{figure}
 \centering
 \includegraphics[width=1\columnwidth]{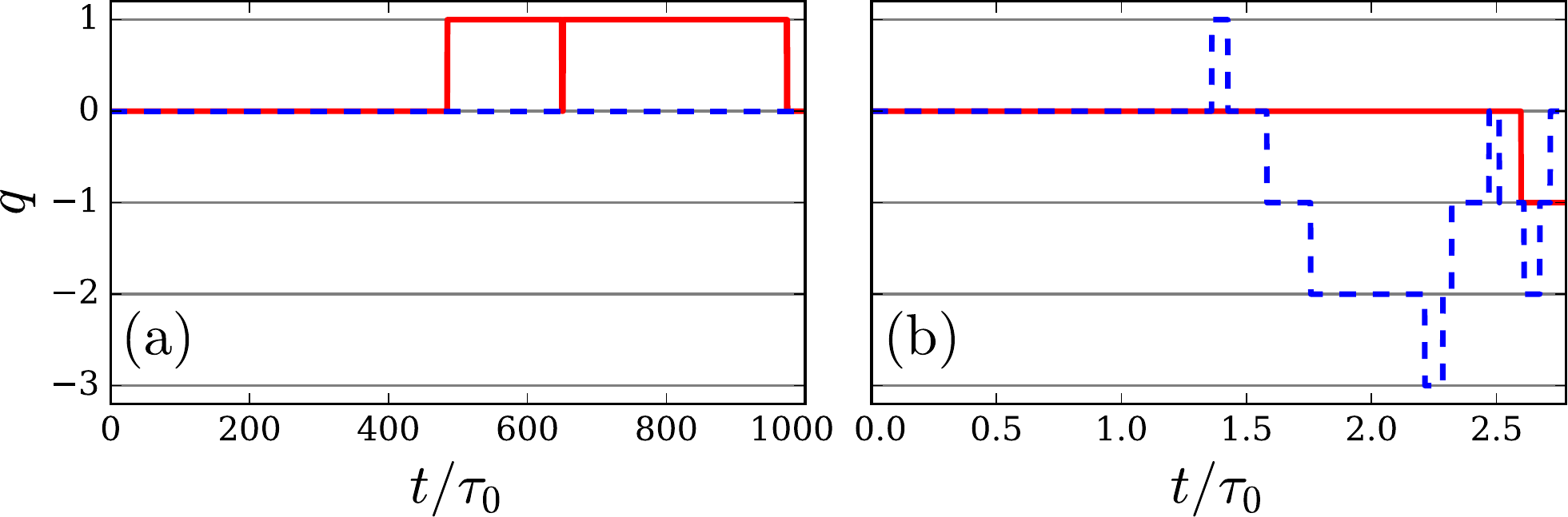}
 \caption{Time evolution of the winding number $q$ for Eq.\eqref{eq:kuramoto} on a single-cycle network with $n=83$ nodes, 
$m=0$ (red lines) and $\frac{m}{d}/\frac{d}{\lambda_2}=10/175$ (blue dashed lines). 
 (a) Noise with short correlation time $\lambda_2\tau_0/d=5.7\cdot 10^{-4}$. (b) Noise with 
 longer correlation time $\lambda_2\tau_0/d=0.03$.}
 \label{fig:K-inertia2}
\end{figure}

For sufficiently weak, bounded noise, fluctuations are small and 
there is no stochastic escape~\cite{Lee18}. Noise makes the system fluctuate
about its equilibrium, and typical deviation amplitudes can be evaluated from a linearized dynamics about the 
equilibrium~\cite{Bam12,Tyl18a,Hae18}. 
The situation becomes fundamentally different for stronger noise. This is illustrated in
Fig.~\ref{fig:K-inertia2}, which shows the time-evolution of the winding number $q$ labelling different equilibrium fixed points of
Kuramoto-like models, Eq.~\eqref{eq:kuramoto}, with additive Ornstein-Uhlenbeck noise. 
Changes in $q$ indicate that the system visits other basins of attraction, surrounding different equilibrium states. 
Depending on the oscillator inertia and the noise amplitude and correlation time, this happens 
more or less quickly and for longer or shorter periods of time.
Due to the high dimensionality of the state space and the nonlinear coupling between oscillators, the exact
shape and size of the basins are impossible to capture~\cite{Wil06,Men13,Del17b}, consequently, the escape time from one basin
is hard to predict. 
For the Kuramoto model with cyclic interactions, DeVille~\cite{Dev12} showed that the escape time scales as the exponential of the 
potential barrier height between the initial and final equilibrium states. In the spirit of Kramers~\cite{Kra40}, Hindes and 
Schwartz~\cite{Hin16,Hin18} further relate the escape time to the numerically computed action on the action-minimizing 
trajectory between the two equilibria. It is hard to see how these numerical approaches could give
analytical estimates for stochastic escape in higher dimension. 

In this manuscript, we follow an altogether different approach. We specify to synchronous fixed points of Kuramoto-like models,
but stress that the approach is applicable to more general systems. We subject the initial, synchronous state to additive
Ornstein-Uhlenbeck noise. Linearizing the dynamics about the synchronous state, we calculate the standard deviation of the 
noise-induced fluctuations about that state. The linearized dynamics is no longer accurate when the standard deviation
exceeds some threshold distance $D_c$. Clearly, $D_c$ is 
bounded from above by the distance $\Delta$ between the stable synchronous state and the closest
saddle point to the next basin of attraction. We postulate that $D_c$ is parametrically proportional to $\Delta$, so that
the breakdown of linear response coincides with the occurence of stochastic escapes. 
This postulate 
allows us to derive a criterion for stochastic escape based on the distance $\Delta$ between the initial stable synchronous
fixed point and the nearest saddle point and not as in Kramers' and other approaches~\cite{Kra40,Dev12,Scha17,Hin16,Hin18}
on their potential height difference. We validate numerically our postulate that $D_c \sim \Delta$ for four, very different
networks and furthermore show it gives precise estimates for the first stochastic escape time. 

\textbf{The Model.}	 
We consider generic,  Kuramoto-like models of nonlinearly coupled oscillators on complex graphs
defined by the differential equations~\cite{Kur75}
\begin{align}\label{eq:kuramoto}
 m \, \ddot{\theta}_i+d \, \dot{\theta}_i &= P_i - \sum_{j}b_{ij}\sin(\theta_i-\theta_j) \, .
\end{align}
Oscillators with inertia $m$ and damping parameter $d$ are described by compact angle coordinates $ \theta_i\in(-\pi , \pi]$ 
and natural frequencies $P_i\in\mathbb{R}$. They 
are located on nodes $i=1,...,n$ of a connected coupling network defined by the adjacency matrix, $b_{ij}\ge 0$. 
Without loss of generality, we consider $\sum_i P_i=0$, which is equivalent to considering the system in a rotating frame,
because Eq.~\eqref{eq:kuramoto} is invariant under
$\theta_i(t) \rightarrow \theta_i(t)+\Omega t$  $P_i \rightarrow P_i+d \, \Omega$.
For bounded distributions 
of natural frequencies on small enough intervals, synchronous states exist with $\dot{\theta}_i\equiv 0$, $\forall i$.

We consider a stable synchronous state 
${\bm \theta}^{(0)}=(\theta_1^{(0)},\ldots ,\theta_n^{(0)}) $ corresponding to natural frequencies $\bm{P}^{(0)}$. 
We subject this state to a time-dependent perturbation $\bm{P}(t) = \bm{P}^{(0)} + \delta \bm{P}(t)$.
Linearizing the dynamics defined by Eq.~(\ref{eq:kuramoto}) with $\bm{\theta}(t) = \bm{\theta}^{(0)} + \delta \bm{\theta}(t)$, 
one obtains  
\begin{align}\label{eq:kuramoto_lin}
 m\delta \ddot{\bm{\theta}}+d\delta \dot{\bm \theta} &\approx \delta {\bm P} - \mathbb{L}(\{ \theta_i^{(0)} \}) \, \delta {\bm \theta} \, ,
\end{align}
with the weighted Laplacian ${\mathbb L}(\{ \theta_i^{(0)} \})$ defined by
\begin{align}\label{eq:laplacian}
{\mathbb L}_{ij} &= 
\left\{ 
\begin{array}{ll}
-b_{ij} \cos(\theta_i^{(0)} - \theta_j^{(0)}) \, , & i \ne j \, , \\
\sum_k b_{ik} \cos(\theta_i^{(0)} - \theta_k^{(0)}) \, , & i=j \, .
\end{array}
\right.
\end{align}
This matrix is positive semidefinite, 
with a single eigenvalue $\lambda_1=0$ and associated eigenvector ${\bf u}_1=(1,1,1,...1)/\sqrt{n}$, while $\lambda_\alpha>0$, $\alpha=2,3,...n$. 


The dynamics of Eq.~\eqref{eq:kuramoto_lin} is characterized by different times scales. The first one characterizes the 
noisy perturbations. We consider spatially uncorrelated noise with 
vanishing average and Ornstein-Uhlenbeck correlator
\begin{eqnarray}\label{eq:noise}
\langle \delta P_i(t)\delta P_j(t') \rangle=\delta_{ij}\delta P_{0}^2\exp[-|t-t'|/\tau_0] \, .
\end{eqnarray}
Thus, the perturbation is characterized by its variance, $\delta P_0^2$ and its correlation time, $\tau_0>0$. The second time scale
is $m/d$. It gives the typical time over which local excitations are damped by $d$, neglecting the network dynamics. 
Finally, one has a set of 
time scales $d/\lambda_\alpha$, $\alpha=2, ... n$, each of them defined by the ratio of the damping parameter and an 
eigenvalue of the Laplacian. For $m/d > d/4 \lambda_\alpha$ these correspond to oscillation time scales of the Laplacian modes, 
while for $m/d < d/4 \lambda_\alpha$ they give network-dynamical corrections to the damping time scale. 
We consider $\tau_0$ as a tunable parameter allowing us to explore different
regimes depending on its relation with  $m/d$ and $d/\lambda_\alpha$.

We measure the
distance between the state of the system and the initial synchronous state
as the square root of the variance
$\langle \delta {\bm \theta}^2(t) \rangle = \sum_i\langle [\delta {\theta_i}(t)-\delta\overline{\theta}(t)]^2 \rangle $ 
with $\delta\overline{\theta}(t)=n^{-1}\sum_i \delta \theta_i(t)$ and brackets indicating an average over
different realizations of noise with the same first two moments. It appropriately 
gives the standard deviation of the angle deviations in the subspace orthogonal to ${\bf u}_1$, because 
displacements in that subspace do not change the state. 
To calculate $\langle \delta {\bm \theta}^2(t) \rangle$, we expand angle deviations over the eigenbasis of
${\mathbb L}$ and solve Eq.~(\ref{eq:kuramoto_lin}) for the coefficients of that expansion (See Supplemental Material~\cite{SM}).
We obtain the long-time limit
\begin{align}\label{eq:variance_gen}
\begin{split}
\lim_{t\rightarrow \infty}\langle \delta {\bm \theta}^2(t) \rangle 
&= \delta P_0^2 \, \sum_{\alpha \ge 2} \frac{\tau_0+m/d}{\lambda_\alpha (\lambda_\alpha\tau_0 + d + m/ \tau_0)} \; .
\end{split}
\end{align}
In the two  limits of long and short $\tau_0$, one has
\begin{align}\label{eq:asymp}
 \lim_{t\rightarrow\infty}\langle \delta {\bm \theta}^2(t) \rangle &\simeq \left\{ 
 \begin{array}{ll}
 \displaystyle \frac{\delta P_0^2 \tau_0}{nd}\Kf_1  \, , & \tau_0 \ll \frac{d}{\lambda_\alpha}\, , \frac{m}{d}\, ,\vspace{2mm} \\ 
 \displaystyle \frac{\delta P_0^2}{n}\Kf_2  \, , & \tau_0 \gg \frac{d}{\lambda_\alpha} \, ,  \frac{m}{d}\, ,
 \end{array}
 \right. 
\end{align}
with $\Kf_p=n\sum_{\alpha \ge 2}\lambda_{\alpha}^{-p}$ \cite{Kle93,Tyl18a}.
Interestingly, none of these asymptotics depend on inertia. 

\begin{figure*}
 \centering
 \includegraphics[width=.9\textwidth]{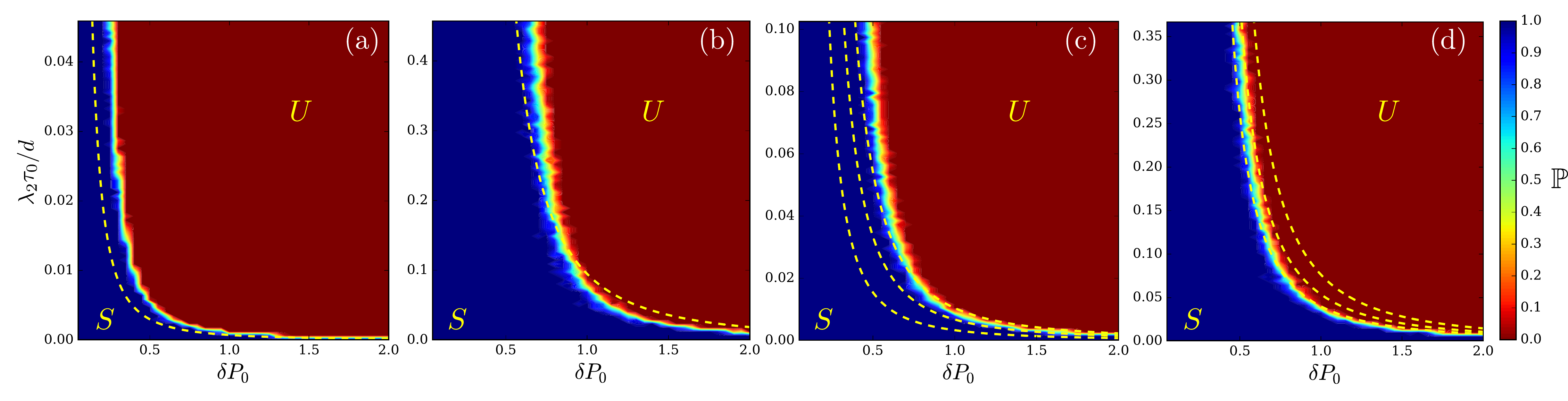}
 \caption{Color-coded survival probability $\mathbb{P}$ for  Eq.~\eqref{eq:kuramoto} with $m=0$. (a) Single-cycle network with $n=83$ and nearest-neighbor coupling; (b) single-cycle network with $n=83$, nearest- and $3^{\mathrm{rd}}$-neighbor coupling; (c) UK transmission network with $n=120$;
 (d) small-world network with $n=200$ nodes (See Supplemental Material~\cite{SM}). 
 Yellow dashed lines are given by Eq.~(\ref{eq:crit}) with $m=0$ and $\Delta$ obtained analytically for panel (a) 
 and numerically for panels (b-d) (See Supplemental Material~\cite{SM}). 
 Observation times $T_{\rm obs}$ correspond to comparable dimensionless parameters $\lambda_2 T_{\rm obs}/d=143$ (a), $143$ (b), $130$ (c) and  $115$ (d).
 }
 \label{fig:escape_simu}
\end{figure*}

\textbf{Escape from the basin.}	
The dynamics of Eq.~\eqref{eq:kuramoto} is described by a vector function
$\bm{\theta}(t)$ following the gradient of the potential 
\begin{align}\label{eq:potential}
 {\cal V}(\bm{\theta},t) &= \sum_{i=1}^nP_i(t)\theta_i - \sum_{i,j}b_{ij}\left[1-\cos(\theta_i-\theta_j)\right]\, ,
\end{align}
starting from $\bm{\theta}(t=0)=\bm{\theta}^{(0)}$.
When the noisy perturbation tilts this potential strongly enough, 
$\bm{\theta}$ can escape the basin of attraction of $\bm{\theta}^{(0)}$.
DeVille showed that, for not too large $\delta P_0$,
the system almost surely escapes the basin in a neighborhood of a $1$-saddle~\cite{Dev12}.
Comparing the typical distance between  $\bm{\theta}$ and $\bm{\theta}^{(0)}$ of Eq.~\eqref{eq:variance_gen} 
with the distance $\Delta$ between $\bm{\theta}^{(0)}$ and its closest $1$-saddle $\bm{\varphi}$
gives us a parametric condition for noise-induced stochastic escape 
\begin{align}\label{eq:crit}
\delta P_{0}^2\sum_{\alpha \ge 2}\frac{\tau_0+m/d}{\lambda_\alpha (\lambda_\alpha\tau_0 + d + m/\tau_0)} &\le \Delta^2 \, .
\end{align}
Our task is therefore to identify the position of the $1$-saddles. This is in general no trivial task because the geometry 
of basins of attraction in such high-dimensional problems is impossible to fully capture. 
For single-cycle networks with identical frequencies, $1$-saddles can be identified analytically~\cite{Dev12,SM}. 
For more general networks, we construct a numerical algorithm 
which locates $1$-saddles $\bm{\varphi}$ and constructs the distribution of their distance to $\bm{\theta}^{(0)}$
(See Supplemental Material~\cite{SM}). 

\textbf{Numerical simulations.}	
We first check  Eq.~\eqref{eq:crit} against numerical simulations of the Kuramoto model of Eq.~\eqref{eq:kuramoto}
with $m=0$. We consider four different networks with constant couplings $b_0=1$ and identical frequencies, which are
a single-cycle network with nearest-neighbor coupling, a single-cycle with nearest- and $3^{\rm rd}$-neighbor coupling, 
a model of the UK transmission network~\cite{Del17b} and a realization of a small-world 
network~\cite{Wat98}. 
Details about these networks are given in the Supplemental Material~\cite{SM}. 
At each node, natural frequencies are perturbed by spatially uncorrelated 
Gaussian noisy sequences $\delta P_i(t)$ satisfying Eq.~\eqref{eq:noise}.
We integrate the dynamics of Eq.~(\ref{eq:kuramoto})  during  an observation time $T_{\rm obs}$ 
and check for a stochastic escape at every time step. Our method for detecting such occurences is based on 
Refs.~\cite{Dor13,Del17a,Man17} which showed 
that on meshed networks, different fixed-point 
solutions of Eq.~\eqref{eq:kuramoto} correspond to a vector of winding numbers ${\bm q}$, each component 
corresponding to one of the cycles of the network.
Refs.~\cite{Dev12,Hin18} observed that transitions between different such equilibrium states occur by phase slips of few
oscillators, and we show in the Supplemental Material~\cite{SM} that these slips can be detected by recording the 
time evolution of ${\bm q}$, as illustrated on Fig.~\ref{fig:K-inertia2}. We therefore detect
desynchronizing events through variations of winding numbers. Details of the method and comments on its accuracy 
are presented in the Supplemental Material~\cite{SM}.
For each set of noise parameters $\delta P_0$ and $\tau_0$
we perform several calculations corresponding to different noise realizations.

Fig.~\ref{fig:escape_simu} shows the 
fraction $\mathbb{P}$ of runs that remain in the initial basin for $t \le T_{\rm obs}$. 
The parameter space is sharply divided into (a) the red region (denoted $U$ for "unstable")
where all runs left the basin of attraction before $T_{\rm obs}$, (b) 
the blue region (denoted $S$ for "stable"), where none of the runs left 
the initial basin of attraction and (c) a rather narrow intermediate region between $U$ and $S$ where some runs left
and some runes stayed in the initial basin. 

It is quite remarkable that the intermediate region (c)
is qualitatively if not quantitatively identifed by Eq.~\eqref{eq:crit} with a network-dependent $\Delta$. As discussed above, 
$\Delta$ is given by a typical distance between the initial stable fixed point $\bm{\theta}^{(0)}$ and 
the nearest saddle point $\bm{\varphi}$ roughly giving the smallest linear size of the basin of attraction. 
For the single-cycle network, all $1$-saddles are located at the same distance from $\bm{\theta}^{(0)}$, which
can be obtained analytically~\cite{Dev12} (See Supplemental Material~\cite{SM}). For the other three networks, 
many, though likely not all 
1-saddles are identified numerically (See the Supplemental Material for details of the method~\cite{SM}). 
For the single-cycle network with nearest- and $3^{\rm rd}$-neighbor coupling, all the $1$-saddles we find are located
at the same distance $\Delta$ from $\bm{\theta}^{(0)}$. 
For the UK and small-world networks, on the other hand, we find a distribution of $\Delta\in [\Delta_{\rm min},\Delta_{\rm max}]$,
which is likely due to the complexity of those meshed networks. The yellow dashed lines in 
Fig.~\ref{fig:escape_simu} then indicate our theoretical prediction Eq.~\eqref{eq:crit} for the obtained value $\Delta$ for the 
two single-cycle networks and for values of $\Delta$ corresponding to the 25$^{\rm th}$, the 50$^{\rm th}$ and 
the 75$^{\rm th}$ precentiles of the distribution of $\Delta$ for the UK and small-world networks. 
In all cases, the shape of the boundary is well predicted. For the more complex UK transmission network, 
Fig.~\ref{fig:escape_simu}(c), there is a horizontal shift between theory and numerics, presumably due to 
to stronger anisotropies of the basins of attraction in this more complex network, effectively 
requiring a larger $T_{\rm obs}$.

In the case of bounded noise, we expect an inertialess system to remain in its initial basin for weak enough noise~\cite{Lee18}.
However, the noise considered in our case is Gaussian and arbitrarily large excursion will occur if one waits long enough. 
As a matter of fact, we found that increasing $T_{\rm obs}$ shifts the boundary between stable and unstable regions
to lower $\delta P_0$ (see Supplemental Material~\cite{SM}). 
Fig.~\ref{fig:fig_esc_final} further shows the stochastic escape time as a function of $\delta P_0$. A superexponential behavior is observed which can be understood as follows.
The noise generates a distribution of angle deviations which we expect to be Gaussian with a variance given 
by Eq.~\eqref{eq:variance_gen}. The escape time is then inversely proportional to the probability to have such a deviation
exceeding $\Delta$, i.e. 
\begin{eqnarray}\label{eq:Tesc}
T_{\mathrm{esc}}\propto \left[2 \int_{\beta\Delta}^\infty P(\overline{\delta \theta})\mathrm{d}(\overline{\delta \theta}) \right]^{-1}
\end{eqnarray}
with a free parameter $\beta$ of order $1$. 
Fig.~\ref{fig:fig_esc_final} validates this argument using a Gaussian distribution of 
single-angle deviation $P(\overline{\delta \theta})$ with variance $\langle \delta {\bm \theta}^2(t) \rangle/n$,
see Eq.~\eqref{eq:variance_gen}. We have found, but do not show, that $T_{\mathrm{esc}}$ diverges at a finite 
value of $\delta P_0$ for a box-distributed, bounded noise.

\begin{figure}
 \centering
 \includegraphics[width=0.8\columnwidth]{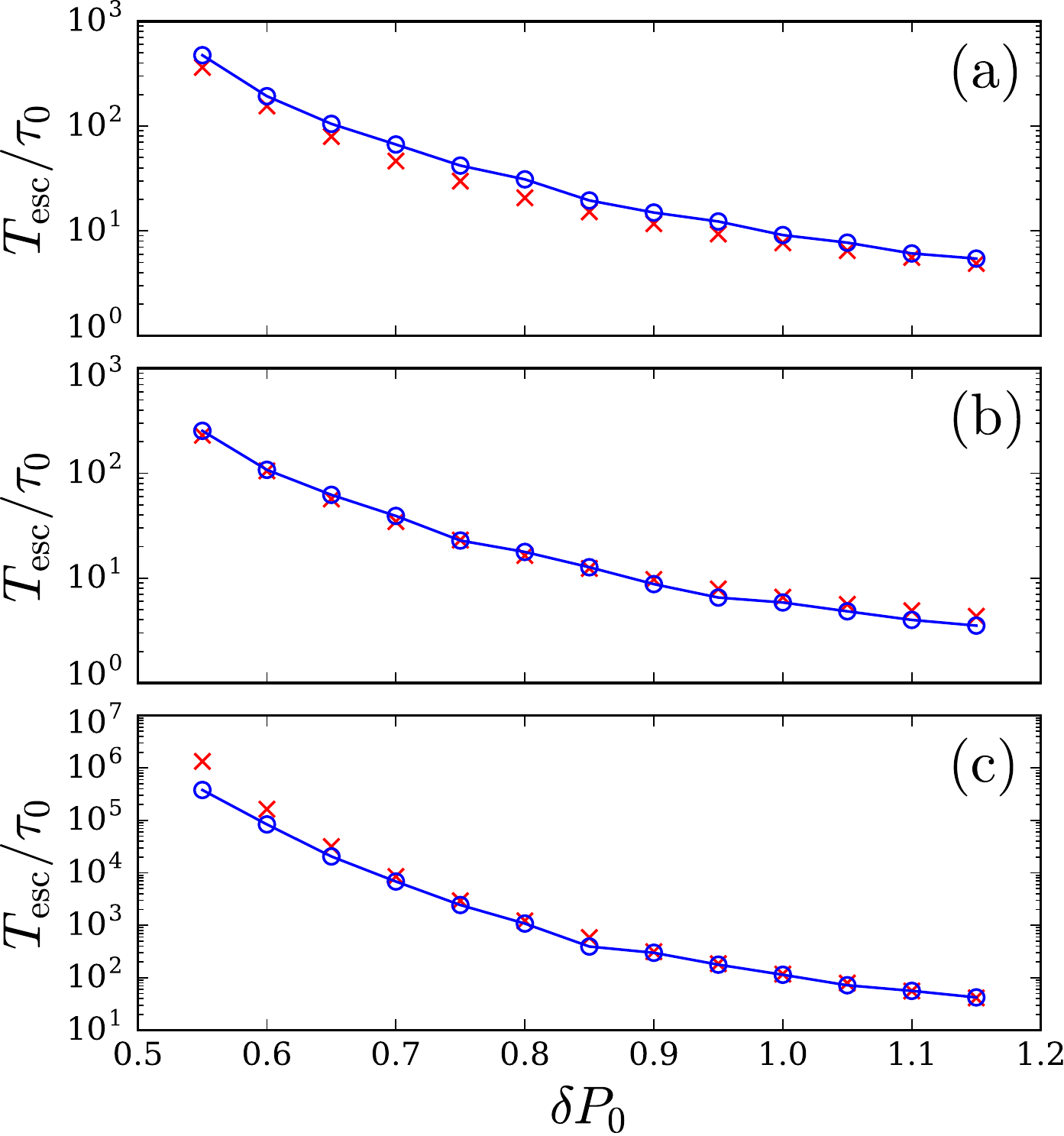}
 \caption{Escape time $T_{\mathrm{esc}}$ from the initial basin of attraction vs. noise amplitude, $\delta P_0$, for cycle networks with $n=83$ (a), $n=249$ (b),  
 and for the UK transmission network (c). The noise correlation time corresponds to $\lambda_2\tau_0/d=8.6\cdot 10^{-3}$ (a), $\lambda_2\tau_0/d=9.6\cdot 10^{-4}$ (b) and $\lambda_2\tau_0/d=0.02$ (c). 
Blue circles are averages over $40$ realizations of noise. Red crosses 
correspond to Eq.~\eqref{eq:Tesc}, with $\beta\cong 5/8$ (a-b) and $\beta\cong 2/5$ (c).}
 \label{fig:fig_esc_final}
\end{figure}

We finally consider Eq.~\eqref{eq:kuramoto} with nonzero inertia. 
We focus on the single-cycle network with nearest- and $3^{\rm rd}$-neighbor coupling, 
and tune the inertia parameter $m$ to explore different regimes defined by the different 
time scales of Eq.~\eqref{eq:kuramoto}. 
Fig.~\ref{fig:K-inertia} shows the difference in survival probabilities with and without inertia
in the regimes (a) ${d}/{\lambda_\alpha}\gtrsim {m}/{d}$, (b) ${d}/{\lambda_\alpha}\lesssim {m}/{d}$ and 
(c) $d/\lambda_\alpha\ll m/d$. 
Deep in the stable (unstable) regions, both inertialess and inertiaful models have $\mathbb P=0$ ($\mathbb P=1$) and 
the difference $\mathbb P(m=0)-\mathbb P(m) = 0$.
Somehow counterintuitively, however, there is an intermediate region where 
the presence of inertia facilitates stochastic escape compared to the inertialess case,
$\mathbb P(m=0)-\mathbb P(m) > 0$. The boundary of that region are 
in excellent agreement with the prediction of Eq.~\eqref{eq:crit}, giving the two dashed yellow lines for $m=0$
and $m\ne0$.

For large $\tau_0$, the faster escape of the system with finite inertia is easily understood. 
With long correlation time, the noise tends to push the system in the same direction for long sequences.
This is sufficient to have the inertiaful system accumulate a significant kinetic energy. The system keeps then moving,
even if, after some time, the noise starts pushing the other way and allows it to move above a saddle point with inertia,
whereas the inertialess system is immediately stopped by noise reversal.

For smaller $\tau_0$, on the other hand, inertia resists short sequences of pushes in rapidly varying directions
and accordingly, we found that inertia stabilizes the system in that case (See Supplemental Material~\cite{SM}). 
This is not predicted by Eq.~\eqref{eq:crit} and is probably due to contributions beyond our linear response theory,
because discrepancies appear for values of $\delta P_0$ comparable to the coupling strength $b_0$. 
The influence of inertia on stochastic escapes is perhaps best illustrated in Fig.~\ref{fig:K-inertia2},
where the presence of inertia stabilizes the system under short-correlated noise [panel (a)] but 
leads to more frequent stochastic escapes for long-correlated noise  [panel (b)].

\begin{figure*}
 \centering
 \includegraphics[width=.9\textwidth]{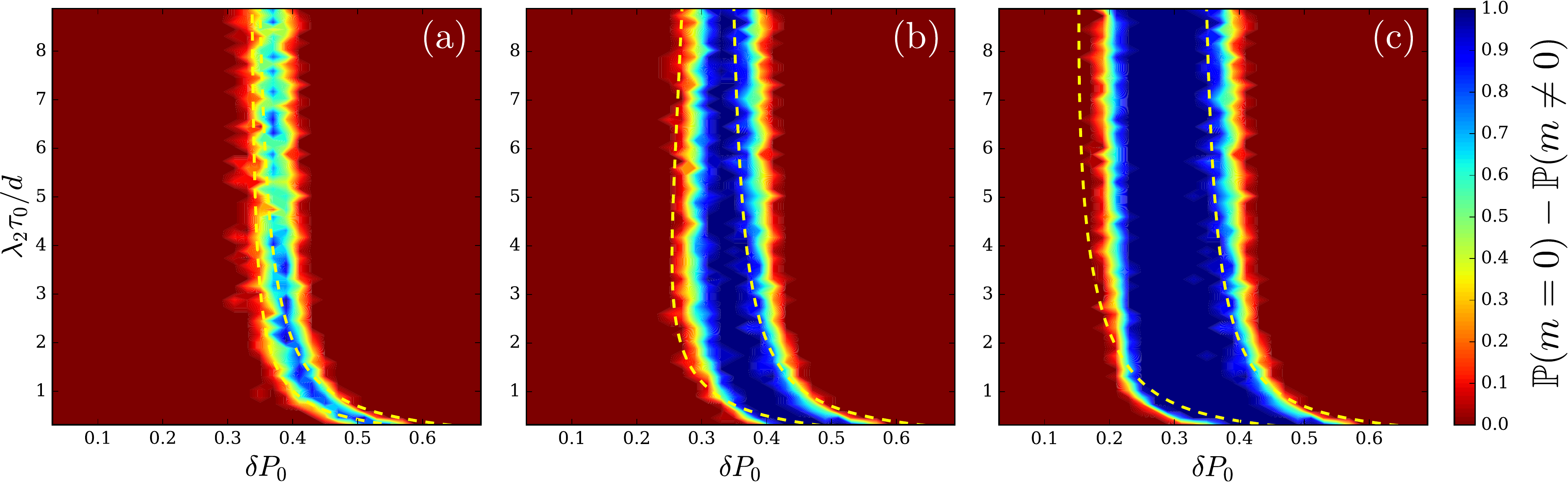}
 \caption{Color-coded difference in escape probability $\mathbb P$ with and without inertia 
 for a single-cycle network with $n=83$ with nearest- and $3^{\rm rd}$-neighbor coupling obtained from $20$ realizations of noise;
 (a) $0.25/0.35$, (b) $2.5/0.35$  and (c) $25/0.35$.  
 The yellow dashed lines are given by Eq.~(\ref{eq:crit}), as discussed in the main text.}
 \label{fig:K-inertia}
\end{figure*}

\textbf{Conclusion.}	 
We have constructed a novel approach to stochastic escape, based on a spectral calculation of typical distances of 
stochastic excursions about equilibrium states and the evaluation 
of  the distance between this equilibrium state and 1-saddles. The method provides analytical results with a single,
model-dependent free parameter of order one [$\beta$ in Eq.~\eqref{eq:Tesc}]. It gives remarkably accurate 
estimates for stochastic escape times, as is illustrated in Fig.~\ref{fig:fig_esc_final}. Interestingly, we found that
the presence of inertia leads to faster, more frequent escapes for long noise coherence times, while the effect is
reversed for short noise coherence times. This is illustrated in Fig.~\ref{fig:K-inertia2}.
Further studies should consider the effect of spatially correlated noise and non-Gaussian, long-tailed noise distributions~\cite{Hae18}.

This work has been supported by the Swiss National Science Foundation under grants 200020\_182050 and 
PYAPP2\_154275.

\pagebreak

\onecolumngrid
\begin{center}
\textbf{\large Noise-Induced Desynchronization and Stochastic Escape from Equilibrium in Complex Networks: Supplemental Material}
\end{center}
\setcounter{equation}{0}
\setcounter{figure}{0}
\setcounter{table}{0}
\makeatletter
\renewcommand{\theequation}{S\arabic{equation}}
\renewcommand{\thefigure}{S\arabic{figure}}
\renewcommand{\bibnumfmt}[1]{[S#1]}
\renewcommand{\citenumfont}[1]{S#1}
\date{\today}

\maketitle

\section{Details of Calculations for the variance of the angle displacements}
We give some details of the calculation that leads to Eq.~(5) of the main text. 
Expanding the angle deviations over the eigenmodes of the Laplacian Eq.~(3) of the main text, 
i.e., $\delta{\bm \theta}(t)=\sum_{\alpha}c_\alpha(t){\bf u}_\alpha$, Eq.~(2) of the main text becomes,
\begin{align}\label{Seq:calpha}
m\,\ddot{c}_\alpha(t)+d\,\dot{c}_\alpha(t) &= \delta {\bm P}(t)\cdot {\bf u}_\alpha - \lambda_\alpha c_\alpha(t)\, , & \alpha &= 2,...,n\, .
\end{align}
With the help of a Laplace transform, the solution of Eq.~\eqref{Seq:calpha} is given by
\begin{align}
c_\alpha(t) &= m^{-1}e^{\frac{-d/m-\Gamma_\alpha}{2}t}\int_0^t e^{\Gamma_\alpha t'}\int_0^{t'} \delta{\bm P}(t'')\cdot {\bf u}_\alpha e^{\frac{d/m - \Gamma_\alpha}{2}t''}dt''dt' \; ,
\end{align}
with $\Gamma_\alpha=\sqrt{(d/m)^2-4\lambda_\alpha/m}$. 
Taking advantage of the orthogonality between eigenmodes of the Laplacian we have,
\begin{align}\label{Seq:var}
\langle\delta {\bm \theta}^2(t)\rangle &\equiv \sum_i\langle [\delta\theta_i(t)-\delta\overline{\theta}(t)]^2\rangle = \sum_{\alpha\ge 2}\langle c_\alpha^2(t)\rangle \;,
\end{align}
with $\delta\overline{\theta}(t)=n^{-1}\sum_i\delta\theta_i(t)$. 
Inserting Eq.~\eqref{Seq:calpha} into Eq.~\eqref{Seq:var}, using the time correlator of $\delta {\bm P}$ Eq.~(4) of the main text, and finally taking the long time limit one obtains, 
after some algebra, Eq.~(5) of the main text.

\section{Method to determine escape time}
Various methods can be used to determine, at any iteration step of the simulation, if the system under consideration has escaped its initial basin of attraction. 
We compared three of them, which we detail here. 

\textbf{Method 1.}
As stated in the main text, stable equilibria of Eq.~\eqref{Seq:dyn} can be unambiguously distinguished by their winding vector $\bm q$. 
The method that we used for the numerical simulations in the main text proceeds as:
\begin{enumerate}
\item At each time step, compute $\bm q$; 
\item If ${\bm q}\neq{\bm q^{(0)}}$ the winding vector of the initial basin of attraction, check if the system is still in the initial basin. 
To do so, simulate the dynamics without noise, taking the current state of the system as initial conditions. 
Once synchrony is reached, compute the winding vector $\bm q^{(1)}$; 
\item If ${\bm q^{(1)}}\neq{\bm q^{(0)}}$, then the system was out of the initial basin. 
Otherwise, if ${\bm q^{(1)}}={\bm q^{(0)}}$, the system was still in the basin and thus the simulation can move to the next time step.
\end{enumerate}

\textbf{Method 2.}
This method is based on DeVille's observation~\cite{Dev12SM} that escapes from basins of attraction occur on a short time interval and can be identified by a fast slip of a small group of angles. 
It proceeds as:  
\begin{enumerate}
\item At each time step, check if some angles made a large excursion, i.e., $\| {\bm \theta}(t)-{\bm \theta^{(0)}} \|_\infty > 2\pi$; 
\item If so, then simulate the dynamics without noise, taking the current state of the system as initial conditions, until it synchronizes to the state ${\bm \theta}^{(1)}$; 
\item If ${\bm \theta^{(1)}}\neq{\bm \theta^{(0)}}$, then the system was out of the initial basin. 
Otherwise, if ${\bm \theta^{(1)}}={\bm \theta^{(0)}}$, the system was still in the basin and thus the simulation can move to the next time step.
\end{enumerate}

\textbf{Method 3.}
Finally, we tested the method in which we check at every time step whether the system returns to the initial basin or not. 
This method guarantees to find the best estimate of the escape time, at least for the Kuramoto model ($m=0$), but is very time-consuming. 

Table \ref{Stab1} compares escape times and final winding numbers for a single-cycle of $n=83$ nodes.
For the Kuramoto model ($m=0$) the three methods give very similar results. 
For the case with inertia, the first two give larger escape times compared to the last method. 
We explain this as follows. 
When the noise is removed, the system may have accumulated some kinetic energy that will drive it out of the basin of attraction. 
And this can happen before the winding number changes or a large angle excursion occurs. 
Furthermore, if the perturbation was still active, it could have pushed the system back towards the stable fixed point before it leaves the basin of attraction, increasing the escape time.
\begin{table}[]
\begin{tabular}{r|ccc|ccc|ccc|ccc|ccc|ccc}
$\rm Simulation$&     & 1 &     &     & 2 &     &     & 3 &     &      & 4 &     &      & 5 &      &      & 6 &  \\
\hline$\rm Method$                        & 1   & 2     & 3   & 1   & 2     & 3   & 1   & 2     & 3   & 1    & 2         & 3   & 1    & 2         & 3    & 1    & 2         & 3    \\
\hline$q^{(1)}$                           & -1  & -1    & -1  & -1  & -1    & -1  & -1  & -1    & -1  & 1    & 1         & -1  & -1   & -1        & -1   & 1    & 1         & -1   \\
\#\rm iterations & 400 & 400   & 400 & 685 & 685   & 685 & 558 & 558   & 550 & 1609 & 1609      & 950 & 1664 & 1664      & 1249 & 1887 & 1887      & 1151    
\end{tabular}
\caption{Final winding number $q^{(1)}$ and number of iterations before the escape for $m=0$ (simulations 1-3) and finite inertia (simulations 4-6). Each triplet is obtained by integrating Eq.~(1) of the main text with the same noise sequence.}\label{Stab1}
\end{table}

\section{The four networks}
We briefly describe the networks used for the numerical simulations of the main text.

\subsection{Cycle with nearest neighbors coupling}
We consider a cycle network of size $n$, with identical natural frequencies. 
The eigenvalues of its weighted Laplacian, Eq.~(3) of the main text, can be obtained analytically,
\begin{align}
\lambda_{\alpha} &= \cos(\delta)[2-2\cos(k_{\alpha})] \, , & \alpha &= 1,...,n \, ,
\end{align}
where $\delta$ is the angle difference between neighboring sites (which are identical at a stable equilibrium~\cite{Del16SM}) and $k_{\alpha}=2\pi(\alpha-1)n^{-1}$. 
For $n=83$ we have $\lambda_{\alpha}\in [0,4\cos(\delta)]$ and $\lambda_2=0.0057$. 

\begin{figure}
 \centering
 \includegraphics[width=.4\textwidth]{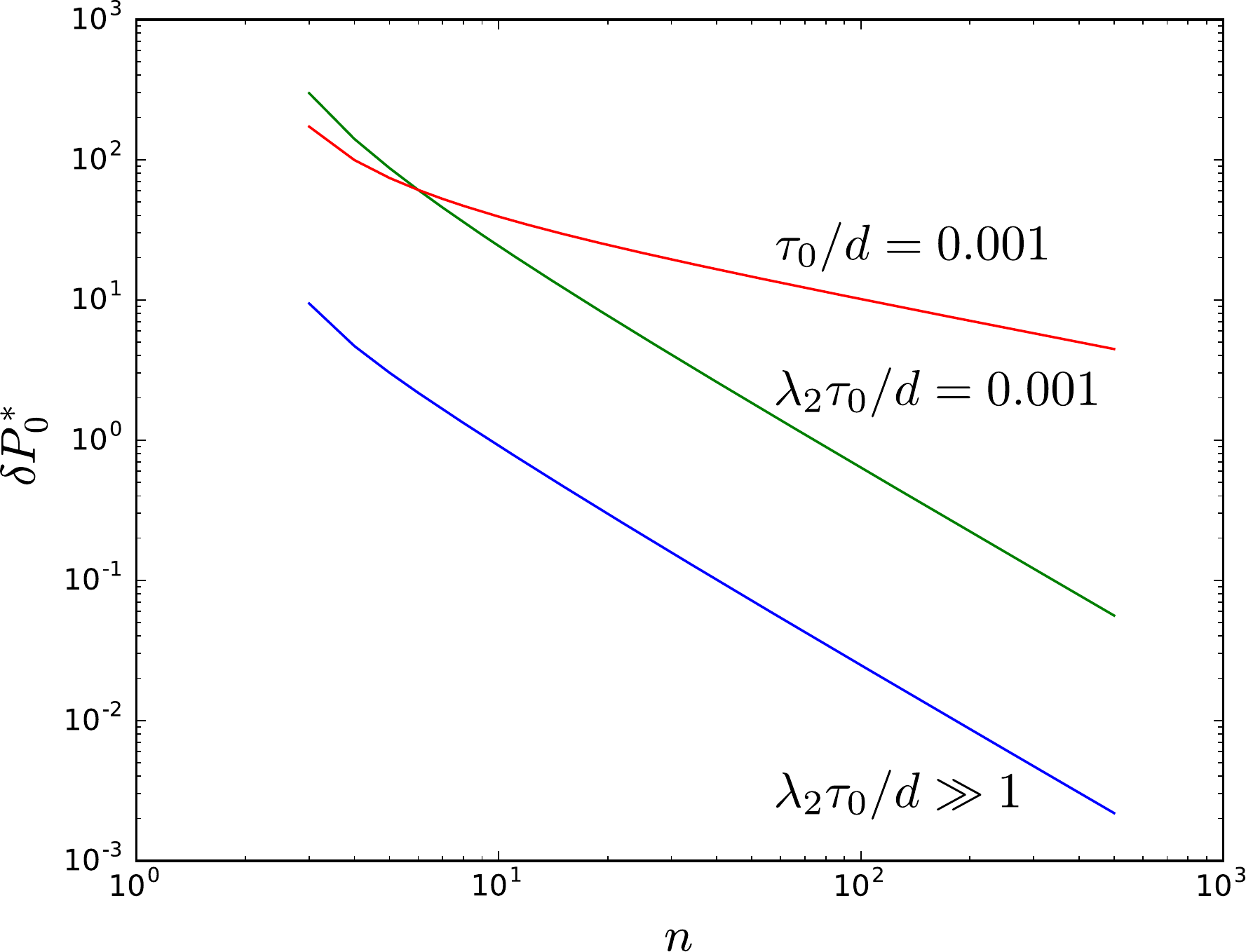}
 \caption{Maximum value $\delta P_0^*$ of the noise amplitude obtained from Eqs.~(\ref{Seq:asymp_cycle1}), (\ref{Seq:asymp_cycle2}) for large (blue) and short (green, red) time correlation, 
 $\tau_0$, as a function of the size of the cyclic network $n$. 
 For the red curve, we consider a constant ratio $\tau_0/d=0.001$. 
 For the green curve we consider a constant ratio $\lambda_2\tau_0/d=0.001$ where $\lambda_2=2-2\cos(2\pi/n)$ depends on the size of the network.}
 \label{Sfig:analytic_cycle}
\end{figure}

Eq.~(6) in the main paper can be explicitly calculated for cyclic networks as functions of the number of nodes $n$ 
\begin{align}\label{Seq:asymp_cycle1}
& \delta P_0^2 \le \frac{\pi^2d n}{\tau_0(n-2)^2}  \, , && \tau_0 \ll d/\lambda_\alpha\, , m/d\, , \\ 
& \delta P_0^2  \le  \frac{60\pi^2n}{(n-2)^2(n^2+11)}  \, , && \tau_0 \gg d/\lambda_\alpha\, , m/d\, .\label{Seq:asymp_cycle2}
\end{align}
Fig.~\ref{Sfig:analytic_cycle} shows the maximum values of $\delta P_0$ satisfying Eqs.~(\ref{Seq:asymp_cycle1}), (\ref{Seq:asymp_cycle2}). 
One remarks that, while increasing the size of the cycle, the stable region gets smaller and even vanishes for $n\rightarrow\infty$ similarly to fluctuations that destroy long-range order in $1$ dimensional locally interacting quantum magnets \cite{Gia04SM}. 

\subsection{Cycle with nearest- and $3^{\mathrm{rd}}$-neighbors coupling}
\begin{figure}
 \centering
 \includegraphics[width=\textwidth]{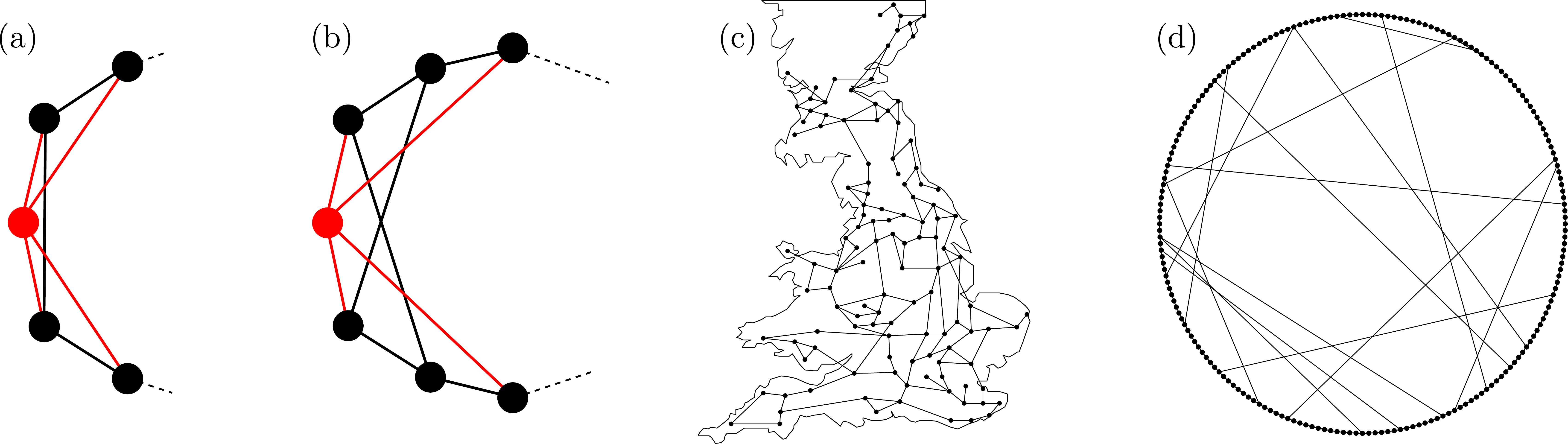}
 \caption{(a) Illustration of the connections of a vertex to its first and second neighbors on a cycle. 
 (b) Illustration of the connections of a vertex to its nearest- and $3^{\rm rd}$-neighbors on a cycle. 
 (c) Illustration of the UK network with $n=120$ vertices and $m=165$ edges. 
 (d) Illustration of our small world network with $n=200$ vertices. 
 Its relative clustering coefficient is $C({\cal G}_p)/C({\cal G}_0)\approx0.89$ and its relative characteristic path length is $L({\cal G}_p)/L({\cal G}_0)\approx0.32$. }
 \label{Sfig:graphs}
\end{figure}
We consider a cycle network of size $n$, where each vertex is connected to its nearest- and $3^{\rm rd}$-neighbors [see Fig.~\ref{Sfig:graphs}(b)]. 
With identical natural frequencies, the eigenvalues of its weighted Laplacian, Eq.~(3) of the main text, can be obtained analytically,
\begin{align}
\lambda_{\alpha} &= \cos(\delta)[4-2\cos(k_{\alpha})-2\cos(3k_{\alpha})] \, , & \alpha &= 1,...,n \, ,
\end{align}
where $\delta$ is the angle difference between neighboring sites (which are identical at a stable steady-state~\cite{Del16SM}) and $k_{\alpha}=2\pi(\alpha-1)n^{-1}$. 
For $n=83$ we have $\lambda_{\alpha}\in [0,8\cos(\delta)]$ and $\lambda_2=0.057$. 

\subsection{UK transmission grid}
Model of the electrical transmission grid of UK depicted in Fig.~\ref{Sfig:graphs}(c). 
It is composed of $120$ nodes and $165$ edges making $44$ cycles. 
During the numerical simulations, to check whether the system has left the initial basin of attraction or not, we check the winding number on each cycle, 
i.e., the winding vector ${\bm q}=(q_1,...,q_{44})$. 
The second eigenvalue of its Laplacian matrix is $\lambda_2\approx0.013$.

\subsection{Small world}
A small world network is constructed from an initial network, where some edges are randomly rewired (see~\cite{Wat98SM}). 
In our case, the initial network ${\cal G}_0$ is a cycle with $n=200$ vertices and where each vertex is connected to its first and second neighbors [see Fig.~\ref{Sfig:graphs}(a)]. 
Each edge $(i,j)$ is then replaced with probability $p=0.05$ by the edge $(i,k)$, where $k$ is chosen at random among the vertices not already connected to $i$. 
The network obtained ${\cal G}_p$ is illustrated in Fig.~\ref{Sfig:graphs}(d). 
It is a small world as it has a large relative clustering coefficient $C({\cal G}_p)/C({\cal G}_0)\approx 0.89$ 
and a small relative characteristic path length $L({\cal G}_p)/L({\cal G}_0)\approx0.32$ 
(see~\cite{Wat98SM} for more details). 
The second eigenvalue of its Laplacian matrix is $\lambda_2\approx0.046$.

\section{Finding 1-saddles}
We detail our methods for finding 1-saddles (equilibria with a unique unstable direction) of the dynamical system 
\begin{align}\label{Seq:dyn}
 m_i\ddot{\theta}_i + d_i\dot{\theta}_i &= P_i^{(0)} + \delta P_i(t) - \sum_jb_{ij}\sin(\theta_i-\theta_j)\, , & i &= 1,...,n\, ,
\end{align}
for arbitrary coupling graph. 

\subsection{Cycle Networks}
For cycle networks with nearest neighbor coupling and identical natural frequencies, the distance between the stable equilibrium $\bm{\theta}^{(0)}=(0,...,0)$, and the 1-saddle $\bm{\varphi}$, 
can be computed analytically as \cite{Del17bSM}
\begin{align}\label{Seq:dist_cycle}
 \Delta^2 &= \left\|\bm{\theta}^{(0)}-\bm{\varphi}\right\|_2^2 = \frac{n(n^2-1)}{12(n-2)^2}\pi^2\, .
\end{align}

\subsection{General Networks}
For general networks, the anisotropy of the basins of attraction renders the 1-saddles complicated to identify analytically. 
We propose a numercial method to locate 1-saddles, which is based on two results of DeVille~\cite{Dev12SM}: 
\begin{itemize}
 \item Escapes from basins of attraction almost always occur in a neighborhood of a 1-saddle of the potential 
 \begin{align}\label{Seq:potential}
  {\cal V}(\bm{\theta}) &= \sum_{i=1}^nP^{(0)}_i\theta_i - \sum_{i<j}b_{ij}\left[1-\cos(\theta_i-\theta_j)\right]\, ;
 \end{align}
 \item Transitions from a basin to another occur on a short time interval compared to the time the system remains in a basin of attraction.
\end{itemize}

\begin{figure}
 \centering
 \includegraphics[width=.6\textwidth]{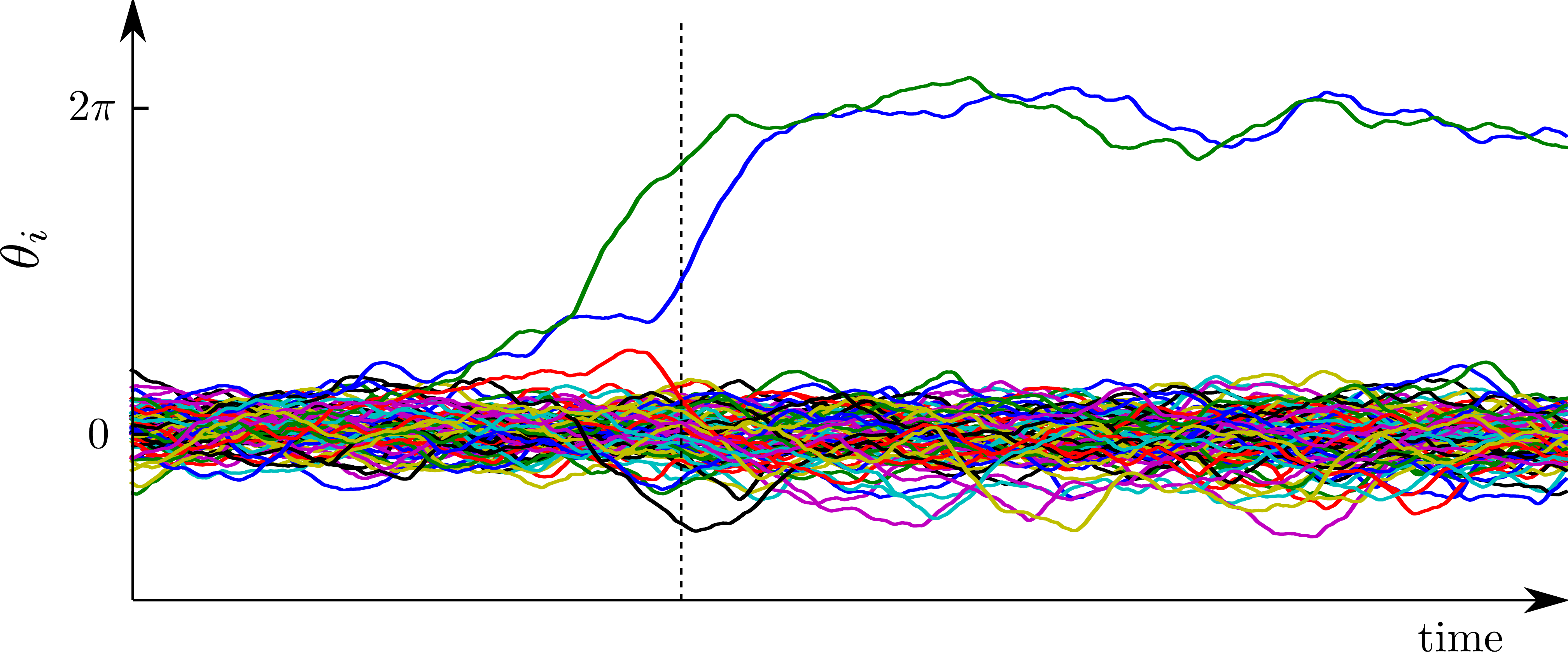}
 \caption{Example of the time evolution of the 120 angles of the UK network [Fig.~\ref{Sfig:graphs}(c)]. 
 We clearly see two angles jumping from a value close to 0 to a value close to $2\pi$. 
 The state of the system at the time given by the vertical dashed line is our candidate for a 1-saddle $\bm{\varphi}$. }
 \label{Sfig:angle_jump}
\end{figure}

We numercially integrate Eq.~\eqref{Seq:dyn}, where $\delta P_i$ is a noise with small variance, 
and keep track of the angles in order to identify iterations where the system is close to a 1-saddle. 
As observed in~\cite{Dev12SM}, when the system is driven (by the noise) to another basin of attraction, its trajectory goes close to a 1-saddle, 
and this can be seen in the time-evolution of the angles as a fast jump of a set of angles of amplitude $2\pi$ (see Fig.~\ref{Sfig:angle_jump}). 
The state $\bm{\varphi}^{(0)}$ of the system in the middle of this jump will be a candidate for a 1-saddle. 
This state is probably not exactly a 1-saddle, but according to~\cite{Dev12SM}, it should be close to one. 
We then solve the steady-state equations 
\begin{align}\label{Seq:steady}
 P_i^{(0)} &= \sum_jb_{ij}\sin(\theta_i-\theta_j)\, , & i &= 1,...,n\, ,
\end{align}
using a Newton-Raphson method with initial conditions $\bm{\varphi}^{(0)}$.
This gives an equilibrium $\bm{\varphi}^*$ of Eq.~\eqref{Seq:dyn}, which we expect to be close to $\bm{\theta}^{(0)}$. 
Computing the eigenvalues of the Jacobian of Eq.~\eqref{Seq:dyn}, the equilibrium $\bm{\varphi}^*$ is a $p$-saddle if and only if it has $p$ positive eigenvalues. 
Note that one eigenvalue is always zero due to invariance of Eqs.~\eqref{Seq:dyn} and \eqref{Seq:potential} under a constant shift of all angles. 

\begin{figure}
 \centering
 \includegraphics[width=.95\textwidth]{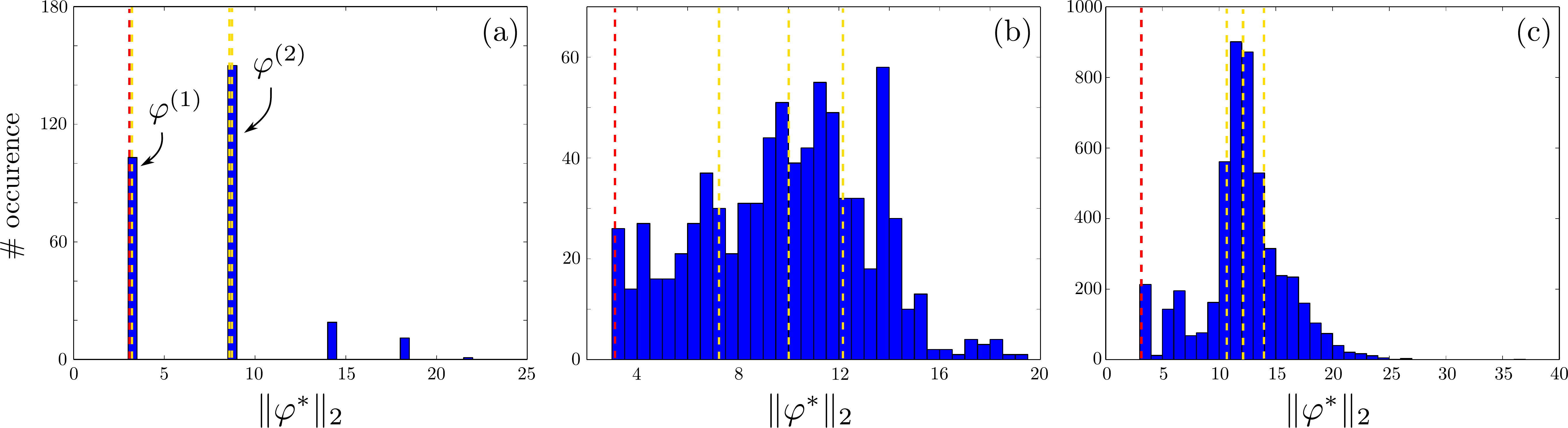}
 \caption{Histograms of the 2-norm distance from the fixed point of the set of 1-saddles found numerically for the cycle with $3^{\rm rd}$-neighbor (a), the UK network (b), 
 and the small world network (c). 
 We found: (a) $284$ 1-saddles for the cycle with $3^{\rm rd}$-neighbor, with smallest 2-norm $n_{\min}\approx 3.12$, and quartiles of the 2-norms $(Q_1,Q_2,Q_3)\approx(3.12,8.61,8.61)$; 
 (b) $788$ 1-saddles for the UK network, with smallest 2-norm $n_{\min}\approx 3.13$, and quartiles of the 2-norms $(Q_1,Q_2,Q_3)\approx(7.24,10.02,12.17)$; and 
 (c) $4956$ 1-saddles for the small-world network, with smallest 2-norm $n_{\min}\approx 3.13$, and quartiles of the 2-norms $(Q_1,Q_2,Q_3)\approx(10.74,12.13,13.95)$. 
 The yellow dashed lines indicate the three quartiles $Q_1$, $Q_2$, and $Q_3$, and the red dashed lines indicate the norm of the closest 1-saddle. }
 \label{Sfig:histograms}
\end{figure}

Running this simulation for a long enough time, we identified: 
\begin{itemize}
 \item $284$ 1-saddles for the cycle with nearest- and $3^{\rm rd}$-neighbor. 
 The distribution of their distance to the stable equilibrium $\bm{\theta}^{(0)}$ is given in Fig.~\ref{Sfig:histograms}(a). 
 Looking more into details, we observe that each value in Fig.~\ref{Sfig:histograms}(a) corresponds to a unique 1-saddle, up to an index shift or the angles' sign reversal. 
 The 1-saddles with the two smallest norm, $\bm{\varphi}^{(1)}$ and $\bm{\varphi}^{(2)}$, are represented in Fig.~\ref{Sfig:phis}. 
 The first one [Fig.~\ref{Sfig:phis}(a)] has the smallest 2-norm, but its configuration with $n-1$ equal angles and one angle $\pi$ apart from all others is, in our opinion, unlikely to occur. 
 As we consider noisy perturbation at all nodes, a configuration with a single large angle excursion and no excursion for all other nodes 
 seems less likely than a configuration where all angles are slightly displaced from their neighbors. 
 In the main text, we performed our study using $\bm{\varphi}^{(2)}$ as 1-saddle for the cycle with nearest- and $3^{\rm rd}$-neighbor. 
 \begin{figure}
  \centering
  \includegraphics[width=.6\textwidth]{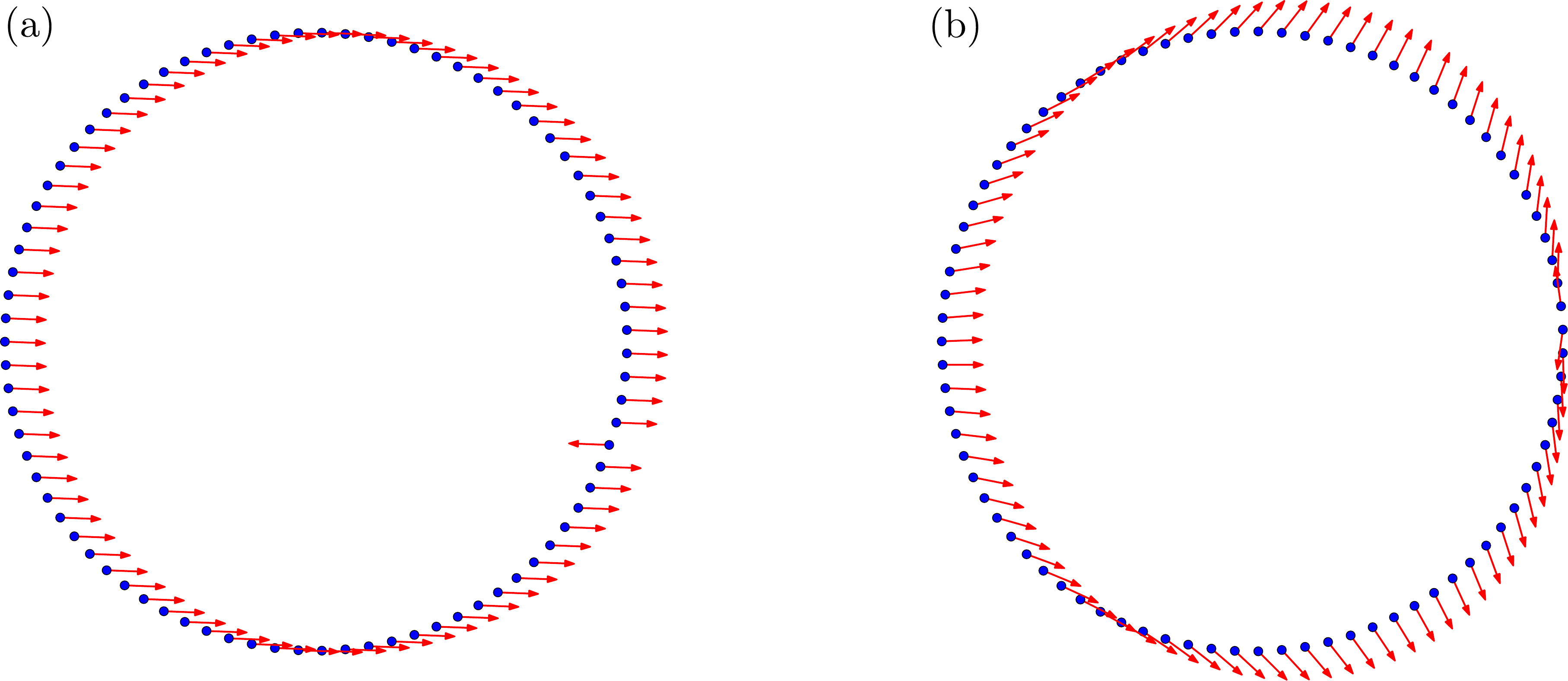}
  \caption{The two 1-saddles, $\bm{\varphi}^{(1)}$ and $\bm{\varphi}^{(2)}$, with smallest 2-norm, for the cycle network, with nearest- and $3^{\rm rd}$-neighbors. 
  (a) $\bm{\varphi}^{(1)}$: all angles are equal, except one which is $\pi$ apart from all others. 
  The 2-norm of this 1-saddles is $\sim 3.12$. 
  (b) $\bm{\varphi}^{(2)}$: all angles are slightly displaced compared to their neighbors. 
  The 2-norm of this 1-saddle is $\sim 8.61$. 
  This configuration is, in our opinion, more likely to occur under noisy perturbations applied to all nodes. }
  \label{Sfig:phis}
 \end{figure}
 \item $788$ 1-saddles for the UK network, whose distribution of the distances to the stable equilibrium is given in Fig.~\ref{Sfig:histograms}(b). 
 Distances cover a large range of value, due to the anisotropy of the basin of attraction; 
 \item $4956$ 1-saddles for the small-world network. 
 The distribution of the distances to $\bm{\theta}^{(0)}$ is given in Fig.~\ref{Sfig:histograms}(c). 
 Most of the 1-saddles are at similar distance. 
\end{itemize}

\section{Superexponential Escape Time}
\begin{figure}
 \centering
 \includegraphics[width=.8\textwidth]{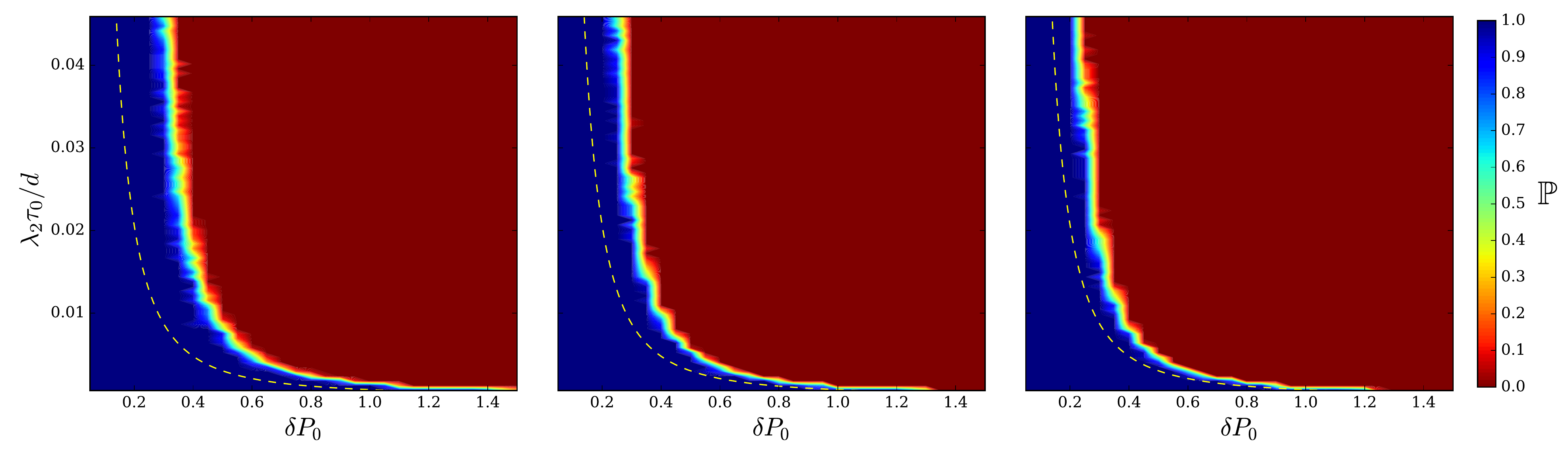}
 \caption{Color plot of the fraction of simulations that stay in the initial basin of attraction obtained from $20$ realizations of Ornstein-Uhlenbeck noisy sequences 
 with amplitude $\delta P_0$ and correlation time $\tau_0$ for a cycle of $n=83$ nodes with $\lambda_2T_{\rm obs}/d=14.3$ (a), $143$ (b), $569$ (c). 
 The yellow dashed line is given by Eq.~(8) of the main text with $m=0$ and $\Delta$ obtained with Eq.~\eqref{Seq:dist_cycle}.}
 \label{Sfig:escape_long_time}
\end{figure}

To evaluate the influence of the observation time $T_{\rm obs}$ on Fig.~(2) of the main text, we performed the simulation for the cycle, increasing the observation time.
Fig.~\ref{Sfig:escape_long_time} shows the fraction on simulations that stay in the initial basin of attraction after an observation time satisfying $\lambda_2T_{\rm obs}/d=14.2$ 
[Fig.~\ref{Sfig:escape_long_time}(a)], $142.4$ [Fig.~\ref{Sfig:escape_long_time}(b)], $569$ [Fig.~\ref{Sfig:escape_long_time}(c)], for a cycle network with $n=83$ nodes. 
As $T_{\rm obs}$ increases exponentially, we observe the boundary between region $U$ and $S$ drifting to the left due to the escape time that is superexponential as $\delta P_0$ decreases.

\section{Linearization Break-Down}
\begin{figure}
 \centering
 \includegraphics[width=0.5\textwidth]{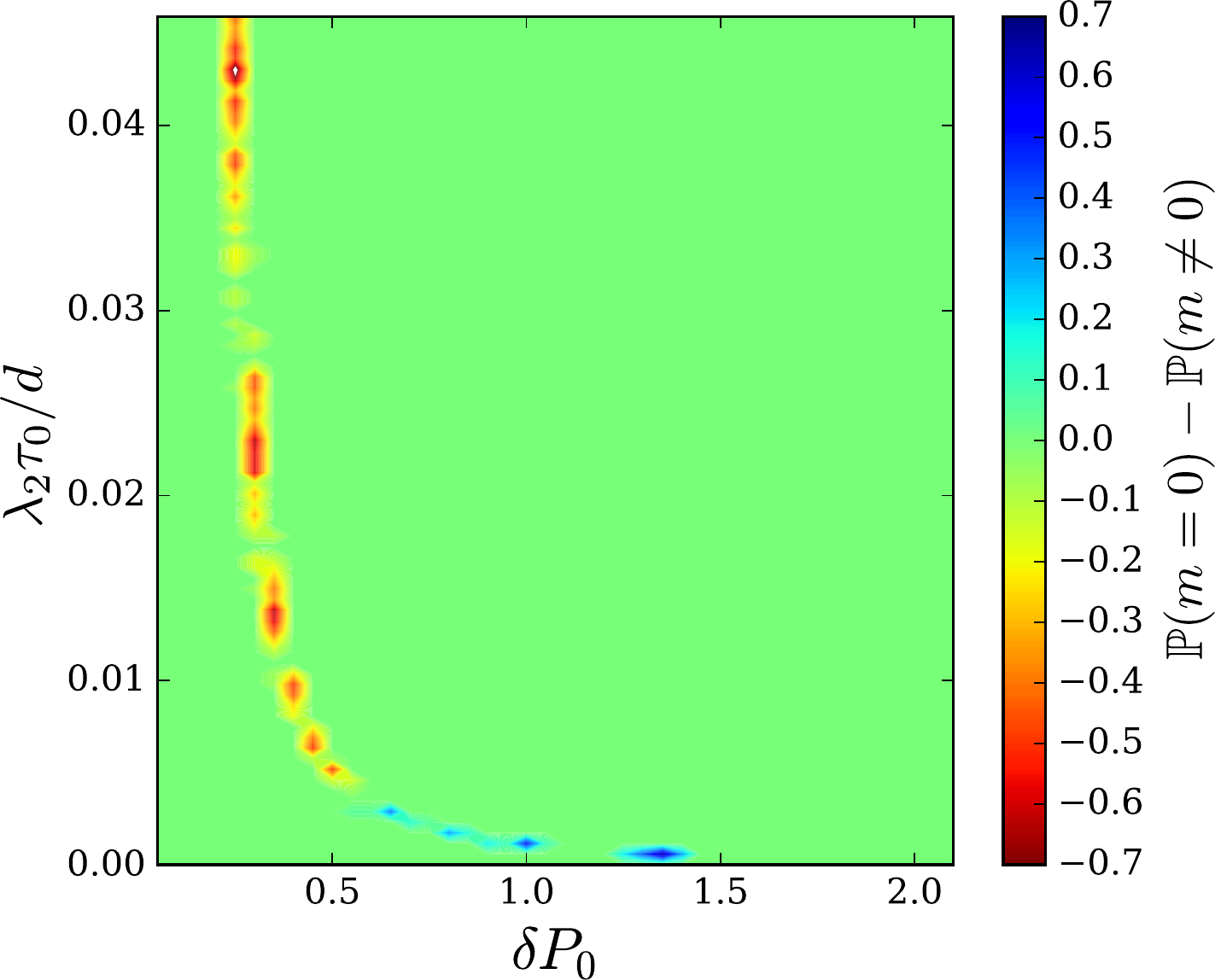}
 \caption{Color plot of the difference of fraction of trajectories that stay in the initial basin of attraction with finite inertia compared to $m=0$ for a cycle network of $n=83$ nodes. 
 Time scales are $\frac{m}{d}\big/\frac{d}{\lambda_2}=10/175$.}
 \label{Sfig:inertia}
\end{figure}

In the main text, we show that, according to our theory, inertia always destabilizes the system compared to the inertialess case. 
However, for the cycle network, we found that for small $\tau_0$ and large $\delta P_0$, inertia stabilizes the system, as illustrated on Fig.~\ref{Sfig:inertia}. 
The blue area where inertia stabilized the system is not predicted by our theory, Eq.~(8) of the main text. 
This can be explained by the breakdown of the linear approximation. 
Indeed, the blue region on Fig.~\ref{Sfig:inertia} starts for value of the order of the coupling $\delta P_0\cong b_0\equiv 1$.


\begin{thebibliography}{29}%
\makeatletter
\providecommand \@ifxundefined [1]{%
 \@ifx{#1\undefined}
}%
\providecommand \@ifnum [1]{%
 \ifnum #1\expandafter \@firstoftwo
 \else \expandafter \@secondoftwo
 \fi
}%
\providecommand \@ifx [1]{%
 \ifx #1\expandafter \@firstoftwo
 \else \expandafter \@secondoftwo
 \fi
}%
\providecommand \natexlab [1]{#1}%
\providecommand \enquote  [1]{``#1''}%
\providecommand \bibnamefont  [1]{#1}%
\providecommand \bibfnamefont [1]{#1}%
\providecommand \citenamefont [1]{#1}%
\providecommand \href@noop [0]{\@secondoftwo}%
\providecommand \href [0]{\begingroup \@sanitize@url \@href}%
\providecommand \@href[1]{\@@startlink{#1}\@@href}%
\providecommand \@@href[1]{\endgroup#1\@@endlink}%
\providecommand \@sanitize@url [0]{\catcode `\\12\catcode `\$12\catcode
  `\&12\catcode `\#12\catcode `\^12\catcode `\_12\catcode `\%12\relax}%
\providecommand \@@startlink[1]{}%
\providecommand \@@endlink[0]{}%
\providecommand \url  [0]{\begingroup\@sanitize@url \@url }%
\providecommand \@url [1]{\endgroup\@href {#1}{\urlprefix }}%
\providecommand \urlprefix  [0]{URL }%
\providecommand \Eprint [0]{\href }%
\providecommand \doibase [0]{http://dx.doi.org/}%
\providecommand \selectlanguage [0]{\@gobble}%
\providecommand \bibinfo  [0]{\@secondoftwo}%
\providecommand \bibfield  [0]{\@secondoftwo}%
\providecommand \translation [1]{[#1]}%
\providecommand \BibitemOpen [0]{}%
\providecommand \bibitemStop [0]{}%
\providecommand \bibitemNoStop [0]{.\EOS\space}%
\providecommand \EOS [0]{\spacefactor3000\relax}%
\providecommand \BibitemShut  [1]{\csname bibitem#1\endcsname}%
\let\auto@bib@innerbib\@empty
\bibitem [{\citenamefont {Ott}(2002)}]{Ott02}%
  \BibitemOpen
  \bibfield  {author} {\bibinfo {author} {\bibfnamefont {E.}~\bibnamefont
  {Ott}},\ }\href {\doibase 10.1017/CBO9780511803260} {\emph {\bibinfo {title}
  {Chaos in Dynamical Systems}}},\ \bibinfo {edition} {2nd}\ ed.\ (\bibinfo
  {publisher} {Cambridge University Press},\ \bibinfo {year}
  {2002})\BibitemShut {NoStop}%
\bibitem [{\citenamefont {van Kampen}(1976)}]{Kam76}%
  \BibitemOpen
  \bibfield  {author} {\bibinfo {author} {\bibfnamefont {N.~G.}\ \bibnamefont
  {van Kampen}},\ }\href {\doibase 10.1016/0370-1573(76)90029-6} {\bibfield
  {journal} {\bibinfo  {journal} {Phys. Rep.}\ }\textbf {\bibinfo {volume}
  {24}},\ \bibinfo {pages} {171} (\bibinfo {year} {1976})}\BibitemShut
  {NoStop}%
\bibitem [{\citenamefont {Machowski}\ \emph {et~al.}(2008)\citenamefont
  {Machowski}, \citenamefont {Bialek},\ and\ \citenamefont {Bumby}}]{Mac08}%
  \BibitemOpen
  \bibfield  {author} {\bibinfo {author} {\bibfnamefont {J.}~\bibnamefont
  {Machowski}}, \bibinfo {author} {\bibfnamefont {J.~W.}\ \bibnamefont
  {Bialek}}, \ and\ \bibinfo {author} {\bibfnamefont {J.~R.}\ \bibnamefont
  {Bumby}},\ }\href@noop {} {\emph {\bibinfo {title} {Power System
  Dynamics}}},\ \bibinfo {edition} {2nd}\ ed.\ (\bibinfo  {publisher} {Wiley},\
  \bibinfo {address} {Chichester, U.K},\ \bibinfo {year} {2008})\BibitemShut
  {NoStop}%
\bibitem [{\citenamefont {Auer}\ \emph {et~al.}(2017)\citenamefont {Auer},
  \citenamefont {Hellmann}, \citenamefont {Krause},\ and\ \citenamefont
  {Kurths}}]{Aue17}%
  \BibitemOpen
  \bibfield  {author} {\bibinfo {author} {\bibfnamefont {S.}~\bibnamefont
  {Auer}}, \bibinfo {author} {\bibfnamefont {F.}~\bibnamefont {Hellmann}},
  \bibinfo {author} {\bibfnamefont {M.}~\bibnamefont {Krause}}, \ and\ \bibinfo
  {author} {\bibfnamefont {J.}~\bibnamefont {Kurths}},\ }\href {\doibase
  10.1063/1.5001818} {\bibfield  {journal} {\bibinfo  {journal} {Chaos}\
  }\textbf {\bibinfo {volume} {27}},\ \bibinfo {pages} {127003} (\bibinfo
  {year} {2017})}\BibitemShut {NoStop}%
\bibitem [{\citenamefont {Sch\"afer}\ \emph {et~al.}(2017)\citenamefont
  {Sch\"afer}, \citenamefont {Matthiae}, \citenamefont {Zhang}, \citenamefont
  {Rohden}, \citenamefont {Timme},\ and\ \citenamefont {Witthaut}}]{Scha17}%
  \BibitemOpen
  \bibfield  {author} {\bibinfo {author} {\bibfnamefont {B.}~\bibnamefont
  {Sch\"afer}}, \bibinfo {author} {\bibfnamefont {M.}~\bibnamefont {Matthiae}},
  \bibinfo {author} {\bibfnamefont {X.}~\bibnamefont {Zhang}}, \bibinfo
  {author} {\bibfnamefont {M.}~\bibnamefont {Rohden}}, \bibinfo {author}
  {\bibfnamefont {M.}~\bibnamefont {Timme}}, \ and\ \bibinfo {author}
  {\bibfnamefont {D.}~\bibnamefont {Witthaut}},\ }\href {\doibase
  10.1103/PhysRevE.95.060203} {\bibfield  {journal} {\bibinfo  {journal} {Phys.
  Rev. E}\ }\textbf {\bibinfo {volume} {95}},\ \bibinfo {pages} {060203(R)}
  (\bibinfo {year} {2017})}\BibitemShut {NoStop}%
\bibitem [{\citenamefont {Gough}\ \emph {et~al.}(1987)\citenamefont {Gough},
  \citenamefont {Colclough}, \citenamefont {Forgan}, \citenamefont {Jordan},
  \citenamefont {Keene}, \citenamefont {Muirhead}, \citenamefont {Rae},
  \citenamefont {Thomas}, \citenamefont {Abell},\ and\ \citenamefont
  {Sutton}}]{Gou87}%
  \BibitemOpen
  \bibfield  {author} {\bibinfo {author} {\bibfnamefont {C.~E.}\ \bibnamefont
  {Gough}}, \bibinfo {author} {\bibfnamefont {M.~S.}\ \bibnamefont
  {Colclough}}, \bibinfo {author} {\bibfnamefont {E.~M.}\ \bibnamefont
  {Forgan}}, \bibinfo {author} {\bibfnamefont {R.~G.}\ \bibnamefont {Jordan}},
  \bibinfo {author} {\bibfnamefont {M.}~\bibnamefont {Keene}}, \bibinfo
  {author} {\bibfnamefont {C.~M.}\ \bibnamefont {Muirhead}}, \bibinfo {author}
  {\bibfnamefont {A.~I.~M.}\ \bibnamefont {Rae}}, \bibinfo {author}
  {\bibfnamefont {N.}~\bibnamefont {Thomas}}, \bibinfo {author} {\bibfnamefont
  {J.~S.}\ \bibnamefont {Abell}}, \ and\ \bibinfo {author} {\bibfnamefont
  {S.}~\bibnamefont {Sutton}},\ }\href {\doibase 10.1038/326855a0} {\bibfield
  {journal} {\bibinfo  {journal} {Nature}\ }\textbf {\bibinfo {volume} {326}},\
  \bibinfo {pages} {855} (\bibinfo {year} {1987})}\BibitemShut {NoStop}%
\bibitem [{\citenamefont {Il'ichev}\ and\ \citenamefont
  {Omelyanchouk}(2008)}]{Ili08}%
  \BibitemOpen
  \bibfield  {author} {\bibinfo {author} {\bibfnamefont {E.}~\bibnamefont
  {Il'ichev}}\ and\ \bibinfo {author} {\bibfnamefont {A.~N.}\ \bibnamefont
  {Omelyanchouk}},\ }\href {\doibase 10.1063/1.2920076} {\bibfield  {journal}
  {\bibinfo  {journal} {Low Temp. Phys.}\ }\textbf {\bibinfo {volume} {34}},\
  \bibinfo {pages} {413} (\bibinfo {year} {2008})}\BibitemShut {NoStop}%
\bibitem [{\citenamefont {Braun}\ \emph {et~al.}(1994)\citenamefont {Braun},
  \citenamefont {Wissing}, \citenamefont {Sch\"afer},\ and\ \citenamefont
  {Hirsch}}]{Bra94}%
  \BibitemOpen
  \bibfield  {author} {\bibinfo {author} {\bibfnamefont {H.~A.}\ \bibnamefont
  {Braun}}, \bibinfo {author} {\bibfnamefont {H.}~\bibnamefont {Wissing}},
  \bibinfo {author} {\bibfnamefont {K.}~\bibnamefont {Sch\"afer}}, \ and\
  \bibinfo {author} {\bibfnamefont {M.~C.}\ \bibnamefont {Hirsch}},\ }\href
  {\doibase 10.1038/367270a0} {\bibfield  {journal} {\bibinfo  {journal}
  {Nature}\ }\textbf {\bibinfo {volume} {367}},\ \bibinfo {pages} {270}
  (\bibinfo {year} {1994})}\BibitemShut {NoStop}%
\bibitem [{\citenamefont {Liu}\ \emph {et~al.}(2018)\citenamefont {Liu},
  \citenamefont {Rui},\ and\ \citenamefont {Duan}}]{Liu18}%
  \BibitemOpen
  \bibfield  {author} {\bibinfo {author} {\bibfnamefont {Y.}~\bibnamefont
  {Liu}}, \bibinfo {author} {\bibfnamefont {C.}~\bibnamefont {Rui}}, \ and\
  \bibinfo {author} {\bibfnamefont {J.}~\bibnamefont {Duan}},\ }\href
  {http://arxiv.org/abs/1811.10960} {\bibfield  {journal} {\bibinfo  {journal}
  {{arXiv:1811.10960}}\ } (\bibinfo {year} {2018})}\BibitemShut {NoStop}%
\bibitem [{\citenamefont {Kramers}(1940)}]{Kra40}%
  \BibitemOpen
  \bibfield  {author} {\bibinfo {author} {\bibfnamefont {H.}~\bibnamefont
  {Kramers}},\ }\href {\doibase https://doi.org/10.1016/S0031-8914(40)90098-2}
  {\bibfield  {journal} {\bibinfo  {journal} {Physica}\ }\textbf {\bibinfo
  {volume} {7}},\ \bibinfo {pages} {284 } (\bibinfo {year} {1940})}\BibitemShut
  {NoStop}%
\bibitem [{\citenamefont {{DeVille}}(2012)}]{Dev12}%
  \BibitemOpen
  \bibfield  {author} {\bibinfo {author} {\bibfnamefont {L.}~\bibnamefont
  {{DeVille}}},\ }\href {\doibase 10.1088/0951-7715/25/5/1473} {\bibfield
  {journal} {\bibinfo  {journal} {Nonlinearity}\ }\textbf {\bibinfo {volume}
  {25}},\ \bibinfo {pages} {1473} (\bibinfo {year} {2012})}\BibitemShut
  {NoStop}%
\bibitem [{\citenamefont {Hindes}\ and\ \citenamefont
  {Schwartz}(2016)}]{Hin16}%
  \BibitemOpen
  \bibfield  {author} {\bibinfo {author} {\bibfnamefont {J.}~\bibnamefont
  {Hindes}}\ and\ \bibinfo {author} {\bibfnamefont {I.~B.}\ \bibnamefont
  {Schwartz}},\ }\href {\doibase 10.1103/PhysRevLett.117.028302} {\bibfield
  {journal} {\bibinfo  {journal} {Phys. Rev. Lett.}\ }\textbf {\bibinfo
  {volume} {117}},\ \bibinfo {pages} {028302} (\bibinfo {year}
  {2016})}\BibitemShut {NoStop}%
\bibitem [{\citenamefont {Hindes}\ and\ \citenamefont
  {Schwartz}(2018)}]{Hin18}%
  \BibitemOpen
  \bibfield  {author} {\bibinfo {author} {\bibfnamefont {J.}~\bibnamefont
  {Hindes}}\ and\ \bibinfo {author} {\bibfnamefont {I.~B.}\ \bibnamefont
  {Schwartz}},\ }\href {\doibase 10.1063/1.5041377} {\bibfield  {journal}
  {\bibinfo  {journal} {Chaos}\ }\textbf {\bibinfo {volume} {28}},\ \bibinfo
  {pages} {071106} (\bibinfo {year} {2018})}\BibitemShut {NoStop}%
\bibitem [{\citenamefont {Lee}\ \emph {et~al.}(2018)\citenamefont {Lee},
  \citenamefont {Aolaritei}, \citenamefont {Vu},\ and\ \citenamefont
  {Turitsyn}}]{Lee18}%
  \BibitemOpen
  \bibfield  {author} {\bibinfo {author} {\bibfnamefont {D.}~\bibnamefont
  {Lee}}, \bibinfo {author} {\bibfnamefont {L.}~\bibnamefont {Aolaritei}},
  \bibinfo {author} {\bibfnamefont {T.~L.}\ \bibnamefont {Vu}}, \ and\ \bibinfo
  {author} {\bibfnamefont {K.}~\bibnamefont {Turitsyn}},\ }\href
  {http://arxiv.org/abs/1803.00817} {\bibfield  {journal} {\bibinfo  {journal}
  {arXiv:1803.00817}\ } (\bibinfo {year} {2018})}\BibitemShut {NoStop}%
\bibitem [{\citenamefont {Bamieh}\ \emph {et~al.}(2012)\citenamefont {Bamieh},
  \citenamefont {Jovanovic}, \citenamefont {Mitra},\ and\ \citenamefont
  {Patterson}}]{Bam12}%
  \BibitemOpen
  \bibfield  {author} {\bibinfo {author} {\bibfnamefont {B.}~\bibnamefont
  {Bamieh}}, \bibinfo {author} {\bibfnamefont {M.~R.}\ \bibnamefont
  {Jovanovic}}, \bibinfo {author} {\bibfnamefont {P.}~\bibnamefont {Mitra}}, \
  and\ \bibinfo {author} {\bibfnamefont {S.}~\bibnamefont {Patterson}},\ }\href
  {\doibase 10.1109/TAC.2012.2202052} {\bibfield  {journal} {\bibinfo
  {journal} {IEEE Trans. Autom. Control}\ }\textbf {\bibinfo {volume} {57}},\
  \bibinfo {pages} {2235} (\bibinfo {year} {2012})}\BibitemShut {NoStop}%
\bibitem [{\citenamefont {Tyloo}\ \emph {et~al.}(2018)\citenamefont {Tyloo},
  \citenamefont {Coletta},\ and\ \citenamefont {Jacquod}}]{Tyl18a}%
  \BibitemOpen
  \bibfield  {author} {\bibinfo {author} {\bibfnamefont {M.}~\bibnamefont
  {Tyloo}}, \bibinfo {author} {\bibfnamefont {T.}~\bibnamefont {Coletta}}, \
  and\ \bibinfo {author} {\bibfnamefont {P.}~\bibnamefont {Jacquod}},\ }\href
  {\doibase 10.1103/PhysRevLett.120.084101} {\bibfield  {journal} {\bibinfo
  {journal} {Phys. Rev. Lett.}\ }\textbf {\bibinfo {volume} {120}},\ \bibinfo
  {pages} {084101} (\bibinfo {year} {2018})}\BibitemShut {NoStop}%
\bibitem [{\citenamefont {Haehne}\ \emph {et~al.}(2018)\citenamefont {Haehne},
  \citenamefont {Schmietendorf}, \citenamefont {Tamrakar}, \citenamefont
  {Peinke},\ and\ \citenamefont {Kettemann}}]{Hae18}%
  \BibitemOpen
  \bibfield  {author} {\bibinfo {author} {\bibfnamefont {H.}~\bibnamefont
  {Haehne}}, \bibinfo {author} {\bibfnamefont {K.}~\bibnamefont
  {Schmietendorf}}, \bibinfo {author} {\bibfnamefont {S.}~\bibnamefont
  {Tamrakar}}, \bibinfo {author} {\bibfnamefont {J.}~\bibnamefont {Peinke}}, \
  and\ \bibinfo {author} {\bibfnamefont {S.}~\bibnamefont {Kettemann}},\ }\href
  {https://arxiv.org/abs/1809.09098v1} {\bibfield  {journal} {\bibinfo
  {journal} {{arXiv}:1809.09098}\ } (\bibinfo {year} {2018})}\BibitemShut
  {NoStop}%
\bibitem [{\citenamefont {Wiley}\ \emph {et~al.}(2006)\citenamefont {Wiley},
  \citenamefont {Strogatz},\ and\ \citenamefont {Girvan}}]{Wil06}%
  \BibitemOpen
  \bibfield  {author} {\bibinfo {author} {\bibfnamefont {D.~A.}\ \bibnamefont
  {Wiley}}, \bibinfo {author} {\bibfnamefont {S.~H.}\ \bibnamefont {Strogatz}},
  \ and\ \bibinfo {author} {\bibfnamefont {M.}~\bibnamefont {Girvan}},\ }\href
  {\doibase 10.1063/1.2165594} {\bibfield  {journal} {\bibinfo  {journal}
  {Chaos}\ }\textbf {\bibinfo {volume} {16}},\ \bibinfo {pages} {015103}
  (\bibinfo {year} {2006})}\BibitemShut {NoStop}%
\bibitem [{\citenamefont {Menck}\ \emph {et~al.}(2013)\citenamefont {Menck},
  \citenamefont {Heitzig}, \citenamefont {Marwan},\ and\ \citenamefont
  {Kurths}}]{Men13}%
  \BibitemOpen
  \bibfield  {author} {\bibinfo {author} {\bibfnamefont {P.~J.}\ \bibnamefont
  {Menck}}, \bibinfo {author} {\bibfnamefont {J.}~\bibnamefont {Heitzig}},
  \bibinfo {author} {\bibfnamefont {N.}~\bibnamefont {Marwan}}, \ and\ \bibinfo
  {author} {\bibfnamefont {J.}~\bibnamefont {Kurths}},\ }\href {\doibase
  10.1038/nphys2516} {\bibfield  {journal} {\bibinfo  {journal} {Nat. Phys.}\
  }\textbf {\bibinfo {volume} {9}},\ \bibinfo {pages} {89} (\bibinfo {year}
  {2013})}\BibitemShut {NoStop}%
\bibitem [{\citenamefont {Delabays}\ \emph
  {et~al.}(2017{\natexlab{a}})\citenamefont {Delabays}, \citenamefont {Tyloo},\
  and\ \citenamefont {Jacquod}}]{Del17b}%
  \BibitemOpen
  \bibfield  {author} {\bibinfo {author} {\bibfnamefont {R.}~\bibnamefont
  {Delabays}}, \bibinfo {author} {\bibfnamefont {M.}~\bibnamefont {Tyloo}}, \
  and\ \bibinfo {author} {\bibfnamefont {P.}~\bibnamefont {Jacquod}},\ }\href
  {\doibase 10.1063/1.4986156} {\bibfield  {journal} {\bibinfo  {journal}
  {Chaos}\ }\textbf {\bibinfo {volume} {27}},\ \bibinfo {pages} {103109}
  (\bibinfo {year} {2017}{\natexlab{a}})}\BibitemShut {NoStop}%
\bibitem [{\citenamefont {Kuramoto}(1975)}]{Kur75}%
  \BibitemOpen
  \bibfield  {author} {\bibinfo {author} {\bibfnamefont {Y.}~\bibnamefont
  {Kuramoto}},\ }in\ \href {\doibase 10.1007/BFb0013365} {\emph {\bibinfo
  {booktitle} {Lecture Notes in Physics {\bf 39}, International Symposium on
  Mathematical Problems in Theoretical Physics}}},\ \bibinfo {editor} {edited
  by\ \bibinfo {editor} {\bibfnamefont {H.}~\bibnamefont {Araki}}}\ (\bibinfo
  {publisher} {Springer},\ \bibinfo {address} {Berlin},\ \bibinfo {year}
  {1975})\BibitemShut {NoStop}%
\bibitem [{SM()}]{SM}%
  \BibitemOpen
  \href@noop {} {\bibinfo  {journal} {See Supplemental Material for detailed
  analytical calculations, informations about the numerical methods and
  additional figures about the effect of observation time and inertia}\
  }\BibitemShut {NoStop}%
\bibitem [{\citenamefont {Klein}\ and\ \citenamefont {Randi\'c}(1993)}]{Kle93}%
  \BibitemOpen
\bibfield  {journal} {  }\bibfield  {author} {\bibinfo {author} {\bibfnamefont
  {D.~J.}\ \bibnamefont {Klein}}\ and\ \bibinfo {author} {\bibfnamefont
  {M.}~\bibnamefont {Randi\'c}},\ }\href
  {http://link.springer.com/article/10.1007/BF01164627} {\bibfield  {journal}
  {\bibinfo  {journal} {J. Math. Chem.}\ }\textbf {\bibinfo {volume} {12}},\
  \bibinfo {pages} {81} (\bibinfo {year} {1993})}\BibitemShut {NoStop}%
\bibitem [{\citenamefont {Watts}\ and\ \citenamefont {Strogatz}(1998)}]{Wat98}%
  \BibitemOpen
  \bibfield  {author} {\bibinfo {author} {\bibfnamefont {D.~J.}\ \bibnamefont
  {Watts}}\ and\ \bibinfo {author} {\bibfnamefont {S.~H.}\ \bibnamefont
  {Strogatz}},\ }\href {\doibase 10.1038/30918} {\bibfield  {journal} {\bibinfo
   {journal} {Nature}\ }\textbf {\bibinfo {volume} {393}},\ \bibinfo {pages}
  {440} (\bibinfo {year} {1998})}\BibitemShut {NoStop}%
\bibitem [{\citenamefont {D\"orfler}\ \emph {et~al.}(2013)\citenamefont
  {D\"orfler}, \citenamefont {Chertkov},\ and\ \citenamefont {Bullo}}]{Dor13}%
  \BibitemOpen
  \bibfield  {author} {\bibinfo {author} {\bibfnamefont {F.}~\bibnamefont
  {D\"orfler}}, \bibinfo {author} {\bibfnamefont {M.}~\bibnamefont {Chertkov}},
  \ and\ \bibinfo {author} {\bibfnamefont {F.}~\bibnamefont {Bullo}},\ }\href
  {\doibase 10.1073/pnas.1212134110} {\bibfield  {journal} {\bibinfo  {journal}
  {Proc. Natl. Acad. Sci.}\ }\textbf {\bibinfo {volume} {110}},\ \bibinfo
  {pages} {2005} (\bibinfo {year} {2013})}\BibitemShut {NoStop}%
\bibitem [{\citenamefont {Delabays}\ \emph
  {et~al.}(2017{\natexlab{b}})\citenamefont {Delabays}, \citenamefont
  {Coletta},\ and\ \citenamefont {Jacquod}}]{Del17a}%
  \BibitemOpen
  \bibfield  {author} {\bibinfo {author} {\bibfnamefont {R.}~\bibnamefont
  {Delabays}}, \bibinfo {author} {\bibfnamefont {T.}~\bibnamefont {Coletta}}, \
  and\ \bibinfo {author} {\bibfnamefont {P.}~\bibnamefont {Jacquod}},\ }\href
  {\doibase 10.1063/1.4978697} {\bibfield  {journal} {\bibinfo  {journal} {J.
  Math. Phys.}\ }\textbf {\bibinfo {volume} {58}},\ \bibinfo {pages} {032703}
  (\bibinfo {year} {2017}{\natexlab{b}})}\BibitemShut {NoStop}%
\bibitem [{\citenamefont {Manik}\ \emph {et~al.}(2017)\citenamefont {Manik},
  \citenamefont {Timme},\ and\ \citenamefont {Witthaut}}]{Man17}%
  \BibitemOpen
  \bibfield  {author} {\bibinfo {author} {\bibfnamefont {D.}~\bibnamefont
  {Manik}}, \bibinfo {author} {\bibfnamefont {M.}~\bibnamefont {Timme}}, \ and\
  \bibinfo {author} {\bibfnamefont {D.}~\bibnamefont {Witthaut}},\ }\href
  {\doibase 10.1063/1.4994177} {\bibfield  {journal} {\bibinfo  {journal}
  {Chaos}\ }\textbf {\bibinfo {volume} {27}},\ \bibinfo {pages} {083123}
  (\bibinfo {year} {2017})}\BibitemShut {NoStop}%
\end{thebibliography}

\begin{thebibliography}{29}%
\makeatletter
\providecommand \@ifxundefined [1]{%
 \@ifx{#1\undefined}
}%
\providecommand \@ifnum [1]{%
 \ifnum #1\expandafter \@firstoftwo
 \else \expandafter \@secondoftwo
 \fi
}%
\providecommand \@ifx [1]{%
 \ifx #1\expandafter \@firstoftwo
 \else \expandafter \@secondoftwo
 \fi
}%
\providecommand \natexlab [1]{#1}%
\providecommand \enquote  [1]{``#1''}%
\providecommand \bibnamefont  [1]{#1}%
\providecommand \bibfnamefont [1]{#1}%
\providecommand \citenamefont [1]{#1}%
\providecommand \href@noop [0]{\@secondoftwo}%
\providecommand \href [0]{\begingroup \@sanitize@url \@href}%
\providecommand \@href[1]{\@@startlink{#1}\@@href}%
\providecommand \@@href[1]{\endgroup#1\@@endlink}%
\providecommand \@sanitize@url [0]{\catcode `\\12\catcode `\$12\catcode
  `\&12\catcode `\#12\catcode `\^12\catcode `\_12\catcode `\%12\relax}%
\providecommand \@@startlink[1]{}%
\providecommand \@@endlink[0]{}%
\providecommand \url  [0]{\begingroup\@sanitize@url \@url }%
\providecommand \@url [1]{\endgroup\@href {#1}{\urlprefix }}%
\providecommand \urlprefix  [0]{URL }%
\providecommand \Eprint [0]{\href }%
\providecommand \doibase [0]{http://dx.doi.org/}%
\providecommand \selectlanguage [0]{\@gobble}%
\providecommand \bibinfo  [0]{\@secondoftwo}%
\providecommand \bibfield  [0]{\@secondoftwo}%
\providecommand \translation [1]{[#1]}%
\providecommand \BibitemOpen [0]{}%
\providecommand \bibitemStop [0]{}%
\providecommand \bibitemNoStop [0]{.\EOS\space}%
\providecommand \EOS [0]{\spacefactor3000\relax}%
\providecommand \BibitemShut  [1]{\csname bibitem#1\endcsname}%
\let\auto@bib@innerbib\@empty
\bibitem [{\citenamefont {{DeVille}}(2012)}]{Dev12SM}%
  \BibitemOpen
  \bibfield  {author} {\bibinfo {author} {\bibfnamefont {L.}~\bibnamefont
  {{DeVille}}},\ }\href {\doibase 10.1088/0951-7715/25/5/1473} {\bibfield
  {journal} {\bibinfo  {journal} {Nonlinearity}\ }\textbf {\bibinfo {volume}
  {25}},\ \bibinfo {pages} {1473} (\bibinfo {year} {2012})}\BibitemShut
  {NoStop}%
\bibitem [{\citenamefont {Delabays}\ \emph {et~al.}(2016)\citenamefont
  {Delabays}, \citenamefont {Coletta},\ and\ \citenamefont {Jacquod}}]{Del16SM}%
  \BibitemOpen
  \bibfield  {author} {\bibinfo {author} {\bibfnamefont {R.}~\bibnamefont
  {Delabays}}, \bibinfo {author} {\bibfnamefont {T.}~\bibnamefont {Coletta}}, \
  and\ \bibinfo {author} {\bibfnamefont {P.}~\bibnamefont {Jacquod}},\ }\href
  {\doibase 10.1063/1.4943296} {\bibfield  {journal} {\bibinfo  {journal} {J.
  Math. Phys.}\ }\textbf {\bibinfo {volume} {57}},\ \bibinfo {pages} {032701}
  (\bibinfo {year} {2016})}\BibitemShut {NoStop}%
\bibitem [{\citenamefont {Giamarchi}(2004)}]{Gia04SM}%
  \BibitemOpen
  \bibfield  {author} {\bibinfo {author} {\bibfnamefont {T.}~\bibnamefont
  {Giamarchi}},\ }\href@noop {} {\emph {\bibinfo {title} {Quantum Physics in
  One Dimension}}}\ (\bibinfo  {publisher} {Oxford University Press},\ \bibinfo
  {year} {2004})\BibitemShut {NoStop}%
\bibitem [{\citenamefont {Watts}\ and\ \citenamefont {Strogatz}(1998)}]{Wat98SM}%
  \BibitemOpen
  \bibfield  {author} {\bibinfo {author} {\bibfnamefont {D.~J.}\ \bibnamefont
  {Watts}}\ and\ \bibinfo {author} {\bibfnamefont {S.~H.}\ \bibnamefont
  {Strogatz}},\ }\href {\doibase 10.1038/30918} {\bibfield  {journal} {\bibinfo
   {journal} {Nature}\ }\textbf {\bibinfo {volume} {393}},\ \bibinfo {pages}
  {440} (\bibinfo {year} {1998})}\BibitemShut {NoStop}%
\bibitem [{\citenamefont {Delabays}\ \emph
  {et~al.}(2017{\natexlab{a}})\citenamefont {Delabays}, \citenamefont {Tyloo},\
  and\ \citenamefont {Jacquod}}]{Del17bSM}%
  \BibitemOpen
  \bibfield  {author} {\bibinfo {author} {\bibfnamefont {R.}~\bibnamefont
  {Delabays}}, \bibinfo {author} {\bibfnamefont {M.}~\bibnamefont {Tyloo}}, \
  and\ \bibinfo {author} {\bibfnamefont {P.}~\bibnamefont {Jacquod}},\ }\href
  {\doibase 10.1063/1.4986156} {\bibfield  {journal} {\bibinfo  {journal}
  {Chaos}\ }\textbf {\bibinfo {volume} {27}},\ \bibinfo {pages} {103109}
  (\bibinfo {year} {2017}{\natexlab{a}})}\BibitemShut {NoStop}%
\end{thebibliography}
\end{document}